\documentclass[aps,prl,showpacs,amsmath,amssymb,amsfonts,floatfix,superscriptaddress,twocolumn,lengthcheck]{revtex4-1}
\usepackage{graphicx}
\usepackage{subfigure}
\usepackage{verbatim}
\usepackage{dcolumn}% Align table columns on decimal point
\usepackage{bm}% bold math
\usepackage{epsf}
\usepackage{xcolor}
\usepackage{hyperref}
\usepackage{hhline}
\usepackage{float}
\usepackage{enumerate}
\usepackage{bbm}
\usepackage{lipsum}
\definecolor{mygreen}{rgb}{0.01, 0.31, 0.59}
\definecolor{myblue}{rgb}{0.01, 0.31, 0.59}
\hypersetup{
	colorlinks=true,   % Options: false (boxed links); true (colored links).
	hypertexnames=false,
	linkcolor=blue,  % Color of internal links.
	citecolor=blue, % Color of links to bibliography.
	urlcolor =blue    % Color of external links.
}

\begin{document}
\title{Long-Time Memory and Ternary Logic Gate Using a Multistable Cavity Magnonic System}

\author{Rui-Chang Shen}
\affiliation{Interdisciplinary Center of Quantum Information, State Key Laboratory of Modern Optical Instrumentation, and Zhejiang Province Key Laboratory of Quantum Technology and Device, Department of Physics, Zhejiang University, Hangzhou 310027, China}

\author{Yi-Pu Wang}
\email{yipuwang@zju.edu.cn}
\affiliation{Interdisciplinary Center of Quantum Information, State Key Laboratory of Modern Optical Instrumentation, and Zhejiang Province Key Laboratory of Quantum Technology and Device, Department of Physics, Zhejiang University, Hangzhou 310027, China}

\author{Jie Li}
\affiliation{Interdisciplinary Center of Quantum Information, State Key Laboratory of Modern Optical Instrumentation, and Zhejiang Province Key Laboratory of Quantum Technology and Device, Department of Physics, Zhejiang University, Hangzhou 310027, China}

\author{Shi-Yao Zhu}
\affiliation{Interdisciplinary Center of Quantum Information, State Key Laboratory of Modern Optical Instrumentation, and Zhejiang Province Key Laboratory of Quantum Technology and Device, Department of Physics, Zhejiang University, Hangzhou 310027, China}

\author{G. S. Agarwal}
\affiliation{Institute for Quantum Science and Engineering and Department of Biological and Agricultural Engineering, and Department of Physics and Astronomy, Texas AM University, College Station, Texas 77843, USA}

\author{J. Q. You}
\email{jqyou@zju.edu.cn}
\affiliation{Interdisciplinary Center of Quantum Information, State Key Laboratory of Modern Optical Instrumentation, and Zhejiang Province Key Laboratory of Quantum Technology and Device, Department of Physics, Zhejiang University, Hangzhou 310027, China}

\date{\today}

\begin{abstract}
Multistability is an extraordinary nonlinear property of dynamical systems and can be explored to implement memory and switches. Here we experimentally realize the tristability in a three-mode cavity magnonic system with Kerr nonlinearity. The three stable states in the tristable region correspond to the stable solutions of the frequency shift of the cavity magnon polariton under specific driving conditions. We find that the system staying in which stable state depends on the history experienced by the system, and this state can be harnessed to store the history information. In our experiment, the memory time can reach as long as 5.11 s. Moreover, we demonstrate the ternary logic gate with good on-off characteristics using this multistable hybrid system. Our new findings pave a way towards cavity magnonics-based information storage and processing.		
\end{abstract}

\maketitle

%In practice, manipulating the interactions in a controlled manner is desired.

\textit{Introduction.---}Light-matter interaction plays a crucial role in information processing, storage, and metrology (see, e.g., Refs.~\onlinecite{RevModPhys.91.025005,FriskKockum2019,Yu2019,Ayuso2019,Rivera2020}). Its utilization allows one to control, configure, and create new phases of matter or light signal. In the past few years, cavity magnonics~\cite{PhysRevLett.111.127003,PhysRevLett.113.083603,PhysRevLett.113.156401,PhysRevApplied.2.054002,PhysRevLett.114.227201,PhysRevB.91.094423,Zhang2015,Tabuchi405,PhysRevLett.117.133602,PhysRevB.94.054433,PhysRevB.93.144420,PhysRevB.93.174427,PhysRevLett.116.223601,PhysRevLett.117.123605,Zhang2017,PhysRevLett.120.057202,PhysRevLett.121.203601,PhysRevB.98.024406,PhysRevLett.121.137203,PhysRevLett.123.127202,PhysRevLett.123.227201,PhysRevB.99.214415,Lachance-Quirion425,PhysRevApplied.13.014053,PhysRevLett.125.147202,PhysRevLett.124.213604,PhysRevB.102.064416,PhysRevLett.125.237201}, which is built on coherently~\cite{PhysRevLett.111.127003,PhysRevLett.113.083603,PhysRevLett.113.156401} or dissipatively~\cite{PhysRevB.98.024406,PhysRevLett.121.137203,PhysRevLett.123.127202,PhysRevLett.123.227201} coupled cavity photons and magnons, has increasingly demonstrated its unique advantages in both fundamental and application researches. The magnon mode can work as a transducer between microwave and optical photons, and the related studies are typically referred to as optomagnonics~\cite{PhysRevB.93.174427,PhysRevLett.116.223601,PhysRevLett.117.123605}. Coherent coupling and entanglement between the magnon mode and superconducting qubits are also experimentally realized~\cite{Tabuchi405,Lachance-Quirione1603150,Lachance-Quirion425}. These achievements show that the cavity magnonics has become a versatile platform for interdisciplinary studies, and also provides a useful building block for hybrid quantum systems~\cite{RevModPhys.85.623,Kurizki3866,Lachance_Quirion_2019,doi:10.1063/5.0020277}.

In ferrimagnetic materials, such as the yttrium iron garnet (YIG), on account of the magnetocrystalline anisotropy, the magnon mode can equivalently act as a nonlinear resonator with Kerr-type nonlinearity~\cite{Gurevich,PhysRevB.94.224410,PhysRevLett.120.057202}. When the magnon mode is driven by a microwave field, the excitation number of magnons increases. The Kerr effect induces a frequency shift and bistability can occur. An intriguing question is whether the extraordinary higher-order multistability~\cite{pisarchik2014control,PhysRevLett.98.236401,Paraso2010,PhysRevLett.122.248301,Jung2014,Iniguez-Rabago2019,PhysRevX.10.021017,Hellmann2020} can be realized in the cavity magnonic system~\cite{PhysRevB.102.104415,PhysRevB.103.104411}. Also, it is known that in a bistable or multistable system, the system staying in which stable state is related to the driven history~\cite{PhysRevLett.93.164501,https://doi.org/10.1002/adfm.200500429,cerna2013ultrafast,PhysRevLett.113.074301},
%so this state will store information about the driven history.
and it is important to characterize this memory effect.

Moreover, logic gates can be implemented by harnessing the stable states as logic states. Most of the current processing systems are based on the binary system, and the information is stored in $``0"$ and $``1"$ bits. The multivalued system can, however, be used to implement ternary, quaternary, or even higher-valued logic gates, significantly reducing the number of devices and the overall system complexity~\cite{https://doi.org/10.1002/advs.202004216}. Furthermore, it has been shown that ternary logic gates working as elementary computing units can be more efficient in artificial intelligence simulations~\cite{kak2018ternary,luo2020mtl,Note1}.

In this Letter, we report the first experimental observation of the multistability in a three-mode cavity magnonic system. This hybrid system is composed of two YIG spheres (1 mm in diameter) strongly coupled to a three-dimensional microwave cavity. Owing to the flexible tunability of the cavity magnonic system, we can continuously tune the system parameters to gradually turn the system from being bistable to multistable. Then we characterize the memory effect of the stable state in the tristable region. We find that the cavity magnonic system exhibits a remarkable long-time memory on the driven history that the system experienced. The memory time can approach several seconds, far exceeding the coherence time of the magnon mode. We further demonstrate the on-off switch of the frequency shifts, which reveals a well-repeatable operability. Finally, we build a ternary logic gate by utilizing three separated stable states as the logic output states, and the inputs correspond to zero, moderate, and high drive powers.

In a prior study on the paramagnetic spin ensemble, bistability with a long relaxation time was also observed at the temperature $\sim 25$~mK~\cite{angerer2017ultralong}, which is achieved via the nonlinearity arising from the high magnon excitations in the magnon-photon coupling. %since considerable magnons can be excited by a drive field for the paramagnetic spin ensemble.
However, one cannot use this nonlinearity for a ferrimagnetic or ferromagnetic spin ensemble, because less magnons can be excited by a drive field before the system enters the unstable regime referred to as the Suhl instability~\cite{Suhl}. Thus, we harness the magnon Kerr effect in the YIG sphere. This involves a different mechanism for the nonlinearity. In addition, a very low temperature is needed for the paramagnetic spin ensemble~\cite{angerer2017ultralong}, while multistability can be implemented at a high temperature in our hybrid system, since YIG has a Curie temperature of 560 K. This makes it possible for the logic and switch devices built on cavity magnonics to work in a wide range of temperatures.

\begin{figure}
\centering
\includegraphics[width=0.49\textwidth]{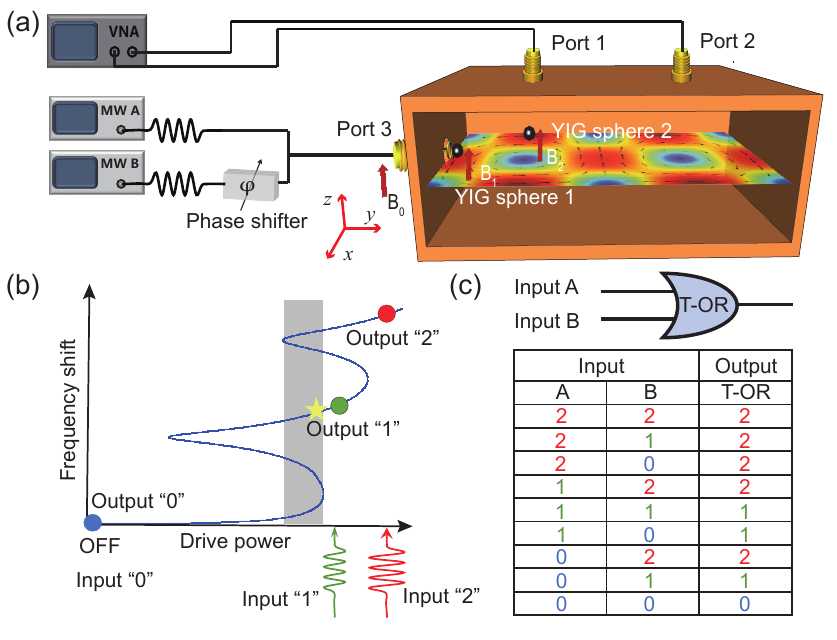}
\caption{(a) Schematic of the three-mode cavity magnonic system, where two YIG spheres are glued on the walls of a microwave cavity. The transmission of the system is measured by the VNA via ports 1 and 2. The drive field is loaded to port 3. The bias magnetic field $\mathbf{B}_{\rm{0}}$ is applied along the \textit{z} direction, and two local magnetic fields $\mathbf{B}_{\rm{1}}$ and $\mathbf{B}_{\rm{2}}$ are applied to separately tune the frequencies of the magnon modes 1 and 2. The inset panel in the cavity shows the magnetic-field distribution of the cavity $\rm{TE}_{102}$ mode. The microwave sources A and B serve as the two inputs of the logic gate. (b) Schematic of the multistability of CMP frequency shift. The gray area is the tristable region. The yellow star marks the state where we measure the memory effect of the multistable system. The blue, green, and red dots correspond to the logic output states 0, 1, and 2 of the ternary $OR$ gate. The corresponding input drive powers are power off (0), 24 dBm (1), 30 dBm (2), respectively.  (c) Symbol and truth table of the ternary $OR$ gate.}
\label{fig1}
\end{figure}

\textit{System and model.---}The experimental setup is schematically shown in Fig.~\ref{fig1}(a). Two 1 mm-diameter YIG spheres are placed in a microwave cavity with a dimension of $44\times20\times6~  {\rm{mm}^{3}}$. The cavity has three ports. Ports 1 and 2 are connected to the vector network analyzer (VNA) for probing the system, and port 3 is terminated with a loop antenna for loading the drive fields. Two microwave sources are combined at port 3 to jointly provide the drive tone, working as input A and input B in the logic gate operation. The YIG spheres are glued on the cavity walls where the magnetic-field antinodes occur for the cavity ${\rm TE}_{\rm{102}}$ mode. Here the YIG sphere 1 is placed near port 3, which can be directly driven by the loop antenna. Because of the antenna-enhanced radiative damping, the damping rate of the Kittel mode~\cite{PhysRev.73.155} in YIG sphere 1 (called magnon model 1) is much larger than that of the Kittel mode in YIG sphere 2 (called magnon model 2). For these two magnon modes and the cavity mode, their damping rates are measured to be $\gamma_{\rm{1}}/2 \pi=13.6~$MHz, $\gamma_{\rm{2}}/2 \pi=1.3~$MHz, and $\kappa_{\rm{c}}/2 \pi=3.2~$MHz. The system is placed in a uniform bias magnetic field ($\mathbf{B}_{\rm{0}}$), and two local magnetic fields ($\mathbf{B}_{\rm{1}}$ and $\mathbf{B}_{\rm{2}}$) are further applied to tune the frequencies of the magnon modes 1 and 2 separately.

When the magnon mode 1 is driven by a microwave field, the hybrid system can be described by
\begin{eqnarray}\label{mul-1}
H/\hbar&=& \omega_{\rm{c}} a^{\dagger} a+\omega_{1} b_{1}^{\dagger} b_{1}+\omega_{2} b_{2}^{\dagger} b_{2}+K_{1} b_{1}^{\dagger} b_{1} b_{1}^{\dagger} b_{1} \nonumber\\
&&+K_{2} b_{2}^{\dagger} b_{2} b_{2}^{\dagger} b_{2}+g_{1}(a^{\dagger} b_{1}+ a b_{1}^{\dagger} )-i g_{2}(a^{\dagger} b_{2}- a b_{2}^{\dagger} ) \nonumber\\
&&-i \Omega_{d}(b_{1} e^{i \omega_{d} t}-b_{1}^{\dagger} e^{-i \omega_{d} t}),
\end{eqnarray}
where $a^{\dag}(a)$ is the creation (annihilation) operator of the cavity photon at frequency $\omega_{\rm{c}}$, $b^{\dag}_{1(2)}(b_{1(2)})$ is the creation (annihilation) operator of the magnon mode 1(2) at frequency $\omega_{1(2)}$, $g_{1(2)}$ is the coupling strength between the magnon mode 1(2) and the cavity mode, $\Omega_{d}$ ($\omega_{d}$) is the drive-field strength (frequency), and $K_{1(2)}$ is the Kerr coefficient of the magnon mode 1(2). When the [100] crystal axis of the YIG sphere is aligned parallel to the bias magnetic field, the Kerr coefficient is positive~\cite{PhysRevLett.120.057202,Gurevich}. This coefficient can be tuned from positive to negative by adjusting the angle between the crystal axis and the bias field~\cite{Gurevich,supp}. It is worth noting that $g_{1}$ and $g_{2}$ have a $\pi/2$ phase difference due to the relative positions of the two YIG spheres in the cavity magnetic field as shown in Fig.~\ref{fig1}(a). We get $g_{1}/2 \pi=41.5~$MHz and $g_{2}/2 \pi=50.5~$MHz by fitting the transmission spectrum.

When a considerable number of magnons are excited, the Kerr term gives rise to a frequency shift of the magnon mode~\cite{supp}: $\chi_{i}=2 K_{i}\langle b_{i}^{\dagger} b_{i}\rangle,~i=1,2$. Using a quantum Langevin approach, we obtain equations for the frequency shifts of the magnon modes and cavity mode~\cite{supp},
\begin{eqnarray}\label{mul-2}
&&[\tilde{\delta}_{\rm{c}}^{2}+(\kappa_{\rm{c}}+\eta_{2} \gamma_{2})^{2}]\Lambda-g_{1}^{2} \chi_{1}=0, \nonumber\\
&&\{(\delta_{1}+\chi_{1}-\eta_{1} \tilde{\delta}_{\rm{c}})^{2}+[\gamma_{1}+\eta_{1}(\kappa_{\rm{c}}+\eta_{2} \gamma_{2})]^{2}\} \chi_{1}=k P_{d}, \nonumber\\
&&[(\delta_{2}+\chi_{2})^{2}+\gamma_{2}^{2}] \chi_{2}-g_{2}^{2}\varepsilon \Lambda=0,
\end{eqnarray}
where $\Lambda=2K_{1}|A|^{2}$, $\varepsilon= K_{2}/K_{1} $, $\delta_{1(2)}=\omega_{\rm{1(2)}}- \omega_{d}$, $\delta_{\rm{c}}=\omega_{\rm{c}}- \omega_{d}$, $\tilde{\delta}_{\rm{c}}=\delta_{\rm{c}}- \eta_{2}(\delta_{2}+\chi_{2})$, $\eta_{2}=g_{2}^{2}/ [(\delta_{2} +\chi_{2})^{2} +\gamma_{2}^{2}]$, $\eta_{1}=g_{1}^{2}/ [\tilde{\delta}_{\rm{c}}^{2}+(\kappa_{\rm{c}}+\eta_{2} \gamma_{2})^{2}]$, and $k P_{d}=2K_{1} | \Omega_{d}|^{2} $. By solving Eq.~(\ref{mul-2}), we can obtain the frequency shifts $\chi_{1}$ and $\chi_{2}$. With these frequency shifts of the individual modes, it is easy to obtain the frequency shifts of the cavity magnon polaritons (CMPs)~\cite{supp}. To find out the multistability, we need to solve the equations numerically to get the CMP frequency shifts at specific drive powers [Fig.~\ref{fig1}(b)]. In this work, we focus on the upper-branch CMP with a frequency of $\omega_{\rm{UP}}$. In the experiment, $K_1/2\pi= 0.0074$~nHz, $K_2/2\pi = 0.0295$~nHz, and the numbers of magnons excited are both less than 3\% of the total number of spins in the two YIG spheres~\cite{supp}.

\begin{figure}[!t]
	\centering
	\includegraphics[width=0.49\textwidth]{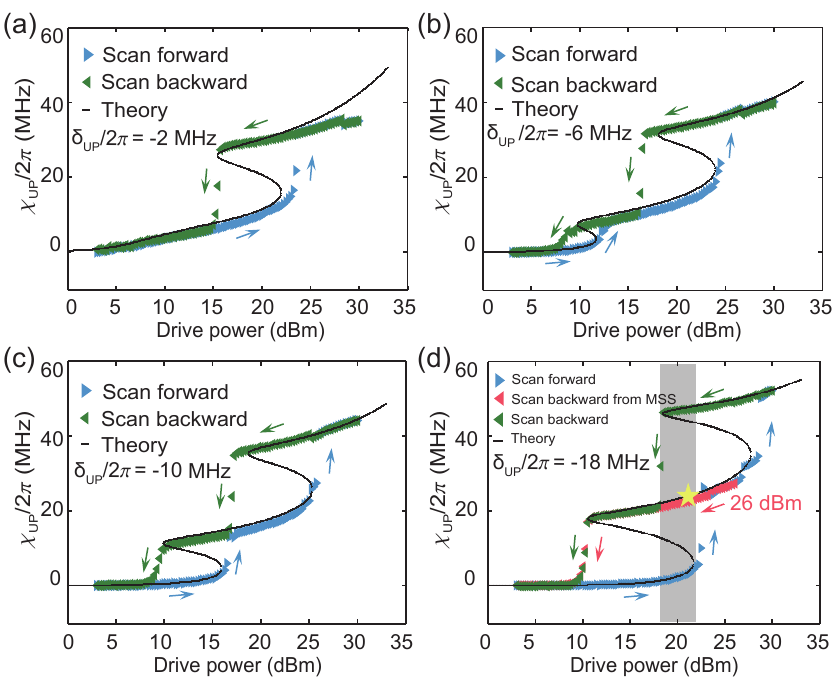}
	\caption{Frequency shift $\chi_{\rm UP}$ of the upper-branch CMP versus the drive power $P_{d}$. The blue (green) triangles are the forward (backward) sweeping results. (a) $\delta_{\rm{UP}}/ 2 \pi\equiv (\omega_{\rm{UP}}-\omega_{d})/2\pi=-2$, (b) $-6$, (c) $-10$, and (d) $-18$~MHz.  The red triangles are obtained by sweeping the drive power backward from 26 dBm, when the upper-branch CMP is in the middle stable state (MSS). The black curves are numerical results calculated using Eq.~(\ref{mul-2}). The gray area depicts the tristable region. The yellow star indicates the state in which we measure the memory effect.}
	\label{fig2}
\end{figure}

\begin{figure}[t]
	\centering
	\includegraphics[width=0.49\textwidth]{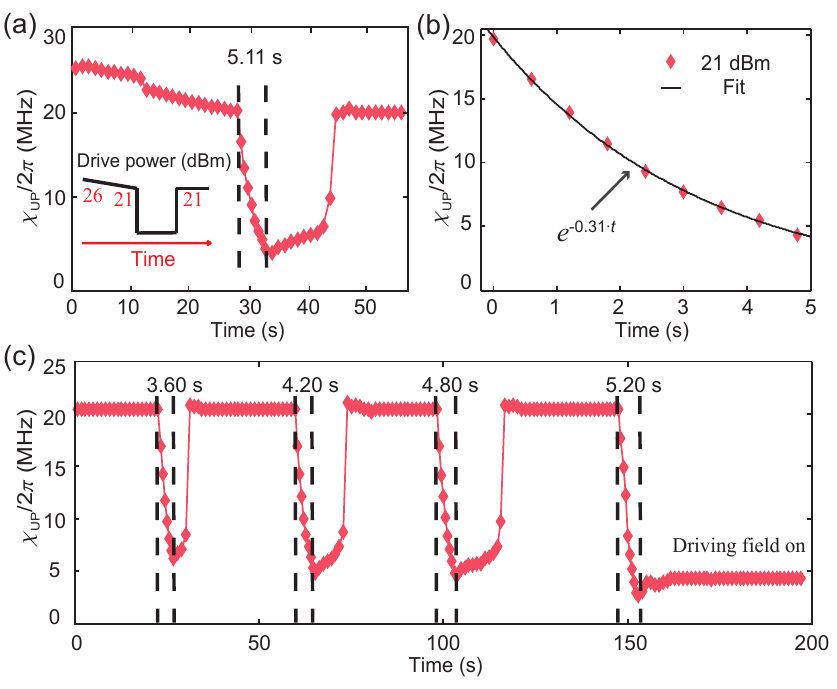}
	\caption{(a) The memory time measurement. The system is first prepared in the MSS located in the tristable region. The drive field is turned off and on to find out the maximum memory time (5.11 s). The inset is the schematic of the drive-power sweeping procedure. (b) The observed frequency-shift relaxation and fitting. The fitting equation is $\chi_{\rm{UP}}/2\pi=19.84\, e^{-\Gamma_{\rm{UP}}t}$~MHz, which gives the relaxation rate $\Gamma_{\rm{UP}}/2\pi=0.049$~Hz. (c) A sequence of switches with different power-off times. From left to right, they are 3.60, 4.20, 4.80, and 5.20 s, respectively. For the power-off time of 5.20 s, we find that the system can no longer recover to the MSS.}
	\label{fig3}
\end{figure}

\textit{Transition from bistability to multistability.---}We tune the two magnon modes to $\omega_{1}/2 \pi=10.212$ and $\omega_{2}/2 \pi=10.198$ GHz, being off resonance to the cavity mode $\rm{TE_{102}}$ ($\omega_{\rm{c}}/2 \pi=10.080$ GHz). In the case with positive Kerr coefficient, the system is found to reach the multistable regime only when the drive-field frequency detuning $\delta_{\rm{UP}}\equiv \omega_{\rm{UP}}-\omega_{d}$ is negative. We monitor the frequency shift $\chi_{\rm{UP}}$ of the upper-branch CMP while sweeping the drive power forward and backward. In Fig.~\ref{fig2}(a), only one hysteresis loop is observed at $\delta_{\rm{UP}}/ 2 \pi=-2$~MHz, corresponding to the bistability of the upper-branch CMP. As we tune $\delta_{\rm{UP}}/ 2 \pi$ to $-6$~MHz, we observe two separated hysteresis loops [Fig.~\ref{fig2}(b)]. In this case, multistability still does not occur. We further tune $\delta_{\rm{UP}}/ 2 \pi$ to $-10$~MHz. Now, two hysteresis loops start to merge with each other, and the multistability is going to emerge [Fig.~\ref{fig2}(c)]. When the detuning becomes $\delta_{\rm{UP}}/ 2 \pi=-$18~MHz, the frequency shift of the upper-branch CMP in Fig.~\ref{fig2}(d) clearly indicates the occurrence of three stable states in the drive-power range from 18.5 to 22 dBm (gray region). The sandwiched state in the tristable region is observed by sweeping the drive power backward when the upper-branch polariton frequency shift is initially prepared in the middle stable state (MSS), as depicted by the red triangles.

Theoretical curves are calculated using Eq.~(\ref{mul-2}), which fit well with the experimental results. Multistability was first predicted for the cavity magnonic system in Ref.~\cite{PhysRevB.102.104415}. Our observations in Fig.~\ref{fig2} convincingly verify this prediction and reveal the excellent tunability of the cavity magnonic system.

\textit{Long-time memory of the stable state.---}Multistability has great potential applications in memory and switches. In previous work, the spin memory of hundreds of exciton polaritons was reported~\cite{cerna2013ultrafast}. In the multistable region, the system staying in which stable state depends on the history the system has experienced. In our system, after turning off the drive field, the higher-energy stable state will have its energy relaxed. Then the frequency shift of the polaritons will quickly vanish. If the microwave drive is turned off for a sufficiently long period, the system can no longer return to the higher-energy stable state when the microwave drive is turned on again. However, there is a {\it critical} time (termed as the memory time) for the relaxed system that can still recover to the higher-energy stable state when the drive field is applied again.
%We term it as the memory time of the system, indicating that the system can recall the history it has experienced within this %time. This memory time is related to the energy relaxation, so it should be limited.
In this work, we experimentally testify that the multistable state of the CMP has a considerably long memory time.

We investigate the memory time of the MSS in the tristable region indicated by the yellow star in Fig.~\ref{fig2}(d). In Fig.~\ref{fig3}(a), we first prepare the CMP in the higher-energy state with a certain frequency shift by applying a 26 dBm drive field. Meanwhile, we turn on the probe field to continuously monitor the frequency shift of the upper-branch CMP. Then we gradually sweep the drive power to 21 dBm. The system now enters the tristable region and reaches the state labeled by the yellow star. This state can be accessed only by performing the aforementioned driving procedure. Therefore, the driven history is stored by the state. Next, we turn off the drive field and keep monitoring the polariton frequency shift. The frequency shift quickly vanishes. After a period of time, we turn on the drive field again, and the system starts to revive and finally returns to the MSS. We term the time interval between turning off and on the drive field as the memory time. In the case of the drive power switching on and off at 21 dBm, the maximum memory time is found to be 5.11~s~\cite{supp}. When the turning-off time exceeds 5.11 s, the CMP frequency shift {\it cannot} recover to the MSS again [see Fig.~\ref{fig3}(c)].

\begin{figure}[!t]
	\centering
	\includegraphics[width=0.48\textwidth]{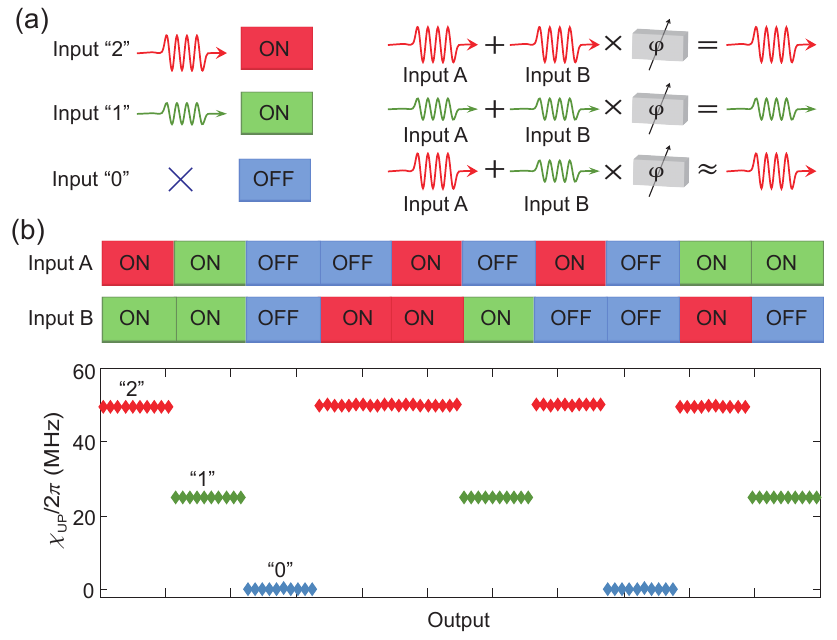}
	\caption{(a) Left panel: Schematic of the input levels. A strong (moderate) drive power represents the 2 (1) input, and the drive power off represents the 0 input. Right panel: The two inputs A and B are combined to drive the upper-branch CMP. To ensure the joint power equal to the larger input of the two inputs, we add a phase shifter in one of the channels and finely adjust the relative phase. (b) Top panel: The sequences of the ternary logic $OR$ gate A-B inputs. Bottom panel: The corresponding outputs of the logic gate. The well-separated output levels 0, 1, and 2 correspond to the zero, small, and large frequency shifts of the upper-branch CMP.  }
	\label{fig4}
\end{figure}

The relaxation rate of the frequency shift reflects the time scale of the magnetocrystalline anisotropy energy relaxation~\cite{supp}. In Fig.~\ref{fig3}(b), the exponential relaxation of the CMP frequency shift is fitted and the relaxation rate is found to be $\Gamma_{\rm{UP}}/2\pi=0.049$~Hz. It shows that this energy relaxation process owns a rate far smaller than the linewidth of the magnon mode. Relaxation rates measured at various initial states are fitted~\cite{supp}, which are all close to $0.049$~Hz.

\textit{Ternary logic gate.---}Our tristable cavity magnonic system is a controllable multivalued system, which can be used to build ternary logic gates. As mentioned above, corresponding to a specific drive power, there is a frequency shift of the upper-branch CMP. The magnitude of these frequency shifts are determined by the theoretical multistability curve solved using Eq.~(\ref{mul-2}). By suitably selecting three input powers, three highly distinguishable frequency shifts can be defined as the $``0,"$ $``1,"$ and $``2"$ output states, as schematically shown in Fig.~\ref{fig1}(b). The corresponding inputs 0, 1, and 2 are illustrated in Fig.~\ref{fig1}(b) and the left panel of Fig.~\ref{fig4}(a). The ternary logic gate we demonstrate is an $OR$ gate, in which, as long as one of the two inputs is 2, the output is 2. If there is no strong drive power (2) input, but there are two or one moderate inputs 1, the output is 1. If both of the two inputs are 0, the output is 0, i.e., there is no CMP frequency shift.

In the case with the two drive fields applied simultaneously, to have the joint drive power equal to the larger input power of the two inputs, we add a phase shifter at one of the input channels. The phase shifter is carefully tuned and fixed to achieve a $2\pi/3$ phase delay between the two input channels~\cite{note2}, which makes the synthesized power of two 2 (1) inputs equal single 2 (1) input [see Fig.~\ref{fig4}(a)]. For a 2 input plus a 1 input, the strong drive power is several times the moderate drive power. After the synthesis, the joint drive power can be approximately equal to the strong drive power. Our follow-up experiments have confirmed this.

In the experiment, we design a series of input A-input B combinations, including all the nine logic configurations listed in the truth table in Fig.~\ref{fig1}(c). The input sequences are shown in the top panel of Fig.~\ref{fig4}(b). The output of the ternary $OR$ gate is displayed in the bottom panel of Fig.~\ref{fig4}(b). We find that different input combinations give excellent distinguishable output states of 0, 1, and 2, which correspond to the zero, small, and large frequency shifts of the upper-branch CMP. These results unambiguously indicate the implementation of an $OR$ gate.

\textit{Conclusions.---}We have experimentally demonstrated the multistability in a three-mode cavity magnonic system. In the tristable region, we find the long-time memory effect of the middle stable state of the CMP frequency shift. The memory time can be as long as 5.11 s, which is millions of times the coherence time of the magnon mode and that of the cavity mode. The switch function of the memory is also characterized within the maximum memory time, which shows a good switchable feature. Utilizing the multivalued cavity magnonic system, we build a ternary logic $OR$ gate, which exhibits highly distinguishable logic states. Our findings offer a novel way towards cavity magnonics-based memory and computing.

\begin{acknowledgments}
This work is supported by the National Natural Science Foundation of China (No.~11934010, No.~U1801661, and No.~11774022), the National Key Research and Development Program of China (No.~2016YFA0301200), Zhejiang Province Program for Science and Technology (No.~2020C01019), and the Fundamental Research Funds for the Central Universities (No.~2021FZZX001-02). G.S.A. thanks AFOSR for support from AFOSR Award No.~FA9550-20-1-0366.
\end{acknowledgments}

\clearpage
\newpage
\onecolumngrid
\begin{center}
	\noindent\textbf{Supplemental Material for\\Long-Time Memory and Ternary Logic Gate Using a Multistable Cavity Magnonic System}
\end{center}
\setcounter{equation}{0}
\setcounter{section}{0}
\setcounter{figure}{0}
\setcounter{page}{1}
\makeatletter
\renewcommand\thesection{S\arabic{section}}	
\renewcommand{\theequation}{S\arabic{equation}}
\renewcommand{\thefigure}{S\arabic{figure}}

\section{\uppercase\expandafter{\romannumeral1}.~The Hamiltonian of the cavity magnonic system}
We study a three-mode cavity magnonic system consisting of two Kittel modes in two YIG spheres and a cavity mode. Among them, one of the magnon modes is driven by a microwave field. The Hamiltonian of the cavity magnonic system is
\begin{equation}\label{sup1}
	H=H _{\rm{c}}+\sum_{i=1}^2(H_{{\rm m},i}+H_{{\rm int},i})+H_{d},
\end{equation}
where $H_{\rm{c}}$ is the Hamiltonian of the cavity mode, $H_{\rm{m,1(2)}}$ is the Hamiltonian of the magnon mode 1(2), $H_{\rm{int,1(2)}}$ is the interaction Hamiltonian between the magnon mode 1 (2) and the cavity mode, and $H_{d}$ is the interaction Hamiltonian related to the drive field. In the experiment, the probe-field power is $-20~\rm{dBm}$, which is much smaller than the drive-field power of $15\sim30~\rm{dBm}$. Also, the probe field is loaded through the cavity port and the drive field is directly applied to the YIG sphere via the antenna, yielding the latter to have a much higher excitation efficiency on the magnons. Therefore, we ignore the magnons excited by the probe field in our experiment.

First, we focus on the Hamiltonian of the magnon mode. Under the bias magnetic field $\mathbf{B}_{0}$, the Hamiltonian reads~\cite{Blundell01, Wang16}
\begin{equation}\tag{S2}\label{sup2}
	H_{\rm{m}}=-\int_{V_{\rm{m}}}\mathbf{M}\cdot\mathbf{B}_{0}d\tau
	-\frac{\mu_{0}}{2}\int_{V_{\rm{m}}}\mathbf{M}\cdot\mathbf{H}_{\rm{an}}d\tau,
\end{equation}
where the first term represents the Zeeman energy and the second term is the magnetocrystalline anisotropy energy. In Eq.~\ref{sup2}, $V_{\rm{m}}$ is the volume of the YIG sphere, $\mathbf{M}=(M_{x},M_{y},M_{z})$ is the magnetization of the YIG sphere, $\mu_{0}$ is the vacuum permeability, and $\mathbf{H}_{\rm{an}}$ is the anisotropic field due to the magnetocrystalline anisotropy in the YIG crystal.

We adopt the direction of the bias magnetic $\mathbf{B}_{0}$ as the $z$ direction  ($\mathbf{B}_{0}=B_{0}\mathbf{e}_{z}$). When the [100] crystal axis of the YIG sphere is aligned along the bias magnetic field, the anisotropic field is given by~\cite{stancil2009spin}
\begin{equation}\label{sup3}
	\mathbf{H}_{\rm{an}}=\frac{2K_{\rm{an}}M_{z}}{\mu_{0}M^{2}}\mathbf{e}_{z},
\end{equation}
where $K_{\rm{an}}$ is the first-order magnetocrystalline anisotropy constant. The Hamiltonian of the magnon mode is
\begin{equation}\label{sup4}
	H_{\rm{m}}=-B_{0}M_{z}V_{\rm{m}}-\frac{K_{\rm{an}}M_{z}^2}{M^{2}}V_{\rm{m}}.
\end{equation}
Using the macrospin operator $\mathbf{S}=\mathbf{M}V_{\rm{m}}/\hbar\gamma\equiv(S_{x},S_{y},S_{z})$~\cite{Soykal10}, where $\gamma$ is the gyromagnetic ratio and $\hbar$ is the reduced Planck constant, the Hamiltonian $H_{\rm{m}}$ can be written as
\begin{equation}\label{sup5}
	\begin{aligned}
		H_{\rm{m}}/\hbar=-\gamma B_{0}S_{z}-\frac{ \hbar K_{\rm{an}}\gamma^2}{M^{2}V_{\rm{m}}}S_{z}^2.
	\end{aligned}
\end{equation}
Here we can define the Kerr coefficient as
\begin{equation}\label{sup6}
	\begin{aligned}
		K=-\frac{\hbar K_{\rm{an}}\gamma^{2}}{M^{2}V_{\rm{m}}}.
	\end{aligned}
\end{equation}
With the parameters $M=143.2~\rm{kA/m}$, $K_{\rm{an}}=-610~\rm{J/m^3}$~\cite{stancil2009spin}, $\gamma/2\pi=28~\rm{GHz/T}$, $\hbar=1.0546\times{10}^{-34}~\rm{J\cdot s}$, and $V_{\rm{m}}=4.19\times{10}^{-9}~\rm{m^{3}}$ for the YIG sphere of 1 mm in diameter, we obtain the Kerr coefficient $K/2 \pi=0.0295~{\rm{nHz}}$ for the [100] crystal axis of the YIG sphere aligned along the bias magnetic field. By rotating the YIG sphere to tune the angle between the crystal axis and the bias magnetic field, the resulting Kerr coefficient can be changed from positive to negative~\cite{Gurevich}.

\begin{figure*}[!htb]
	\centering
	\includegraphics{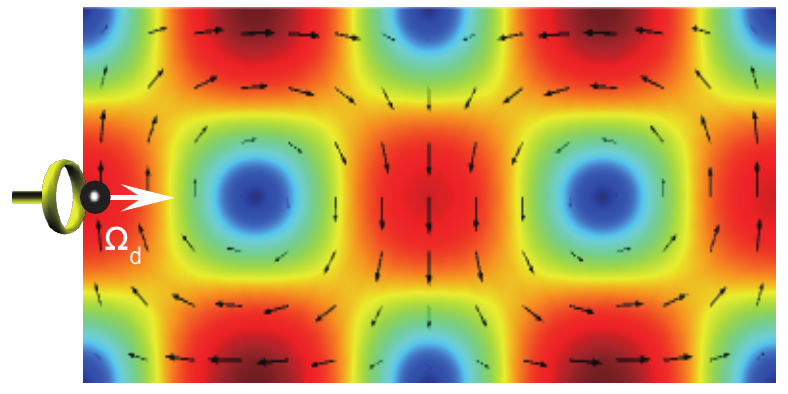}
	\caption{Schematic of the loop antenna for pumping the adjoining YIG sphere and the cavity magnetic field distribution.}
	\label{Fig:S0}
\end{figure*}

To achieve the strong-coupling regime, we place the two YIG spheres at the antinodes of the cavity magnetic field $ \mathbf{h}_{c}$ (cf. Fig.~1 in the main text). Compared with the cavity volume, the YIG spheres are relatively small, so we can assume that the cavity magnetic field is nearly uniform throughout each YIG sphere. Considering the different phases of the cavity magnetic field at the two YIG spheres, we can write the interaction Hamiltonians between the macrospins and the cavity mode as~\cite{Wang16, Soykal10}
\begin{equation}\label{sup7}
	\begin{aligned}
		H_{\rm{int,1}}/\hbar &=2g_{\rm{s,1}} S_{x,1}(a^{\dagger}+a) \\
		&=g_{\rm{s,1}}(S^{+}_{1}+S^{-}_{1})(a^{\dagger}+a),
	\end{aligned}
\end{equation}
\begin{equation}\label{sup8}
	\begin{aligned}
		H_{\rm{int,2}}/\hbar &=2g_{\rm{s,2}} S_{y,2}(a^{\dagger}+a) \\
		&=-ig_{\rm{s,2}}(S^{+}_{2}-S^{-}_{2})(a^{\dagger}+a),
	\end{aligned}
\end{equation}
where the macrospin of the YIG sphere 1(2) couples to the cavity magnetic field along the $x$ ($y$) direction, and $g_{\rm{s,1(2)}}$ is the coupling strength between the macrospin 1(2) and the cavity mode. In the experiment, a drive field of frequency $\omega_{d}$ is applied to directly pump the YIG sphere 1 along the $y$ direction. The corresponding interaction Hamiltonian is
\begin{equation}\label{sup9}
	\begin{aligned}
		H_{d}/\hbar=-i\Omega_{\rm{s}}(S^{+}_{1}-S^{-}_{1})(e^{i \omega_{d} t}+e^{-i \omega_{d} t}),
	\end{aligned}
\end{equation}
where $\Omega_{\rm{s}}$ denotes the coulping strength between the drive field and the macrospin of the YIG sphere 1. The drive magnetic field at port 3 is designed to be perpendicular to the cavity magnetic field. Also, the cavity microwave magnetic field will not pass through the loop of the antenna, which results in a nearly zero coupling between the antenna and the cavity, as shown in Fig.~\ref{Fig:S0}. Therefore, we neglect the drive-field effect on the cavity mode in the Hamiltonian.

Then, we harness the Holstein-Primakoff transformation~\cite{Holstein40}: $S^{+}=\sqrt{2 S-b^{\dagger} b}b$, $S^{-}=b^{\dagger} \sqrt{2 S-b^{\dagger} b}$, and $S_{z}=S-b^{\dag}b$, where $S$ is the total spin of the sample.
The mean excitation numbers of magnons in two YIG spheres can be calculated using the frequency shift $\chi_{1(2)}$ and Kerr coefficient $K_{1(2)}$ via the relation $\chi_{1(2)}=2K_{1(2)}\langle b_{1(2)}^{\dagger} b_{1(2)}\rangle$ (see Eqs.~\ref{sup13}-\ref{sup18} for details). In our experiment, when the drive power is 30 dBm, $\langle b_{1}^{\dagger} b_{1} \rangle = 3.26\times 10^{17}$ and $\langle b_{2}^{\dagger} b_{2} \rangle = 9.45\times 10^{16}$. The excitation numbers are much smaller than the total spin $S$ of a 1 mm-diameter YIG sphere, $S=5.52\times 10^{18}$, calculated using $S=\rho V_{\rm{m}} s$, with $\rho=4.22\times 10^{27}~\rm{m^{-3}}$, $V_{\rm{m}}=\frac{4}{3}\pi R^3$, $R=1$~mm, and $s=\frac{5}{2}$~\cite{tabuchi2014hybridizing,zhang2014strongly}. Therefore, we get $\langle b_{1}^{\dagger} b_{1} \rangle/2S=2.9\%$ and $\langle b_{2}^{\dagger} b_{2} \rangle/2S=0.86\%$. The proportions of the magnons excited with respect to the total spin number are both less than 3$\%$, and the higher-order terms can be neglected in the Holstein-Primakoff transformation. In this case, we have $S^{+}\approx b\sqrt{2S}$ and $S^{-}\approx b^{\dag}\sqrt{2S}$. Under the rotating-wave approximation (i.e., neglecting the fast oscillating terms), the total Hamiltonian of the cavity magnonic system becomes
\begin{eqnarray}\label{sup10}
	\begin{aligned}
		H/\hbar =&\omega_{\rm{c}} a^{\dagger} a+\omega_{1} b_{1}^{\dagger} b_{1}+\omega_{2} b_{2}^{\dagger} b_{2}+K_{1} b_{1}^{\dagger} b_{1} b_{1}^{\dagger} b_{1}+K_{2} b_{2}^{\dagger} b_{2} b_{2}^{\dagger} b_{2} \\
		&+g_{1}(a^{\dagger} b_{1}+a b_{1}^{\dagger})-i g_{2}(a^{\dagger} b_{2}-a b_{2}^{\dagger})-i \Omega_{d}(b_{1} e^{i \omega_{d} t}-b_{1}^{\dagger} e^{-i \omega_{d} t}),
	\end{aligned}
\end{eqnarray}
where $\omega_{1(2)}$ is the frequency of the magnon mode 1(2), $g_{1(2)}=\sqrt{2 S} g_{s,1(2)}$ is the coupling strength between the magnon mode 1(2) and the cavity mode, and $\Omega_{d}=\sqrt{2S} \Omega_{\rm{s}}$ is the drive-field Rabi frequency.
In order to obtain a large multistable region at a low drive power, numerical simulations indicate that it is preferable to choose $K_2$ four times of $K_1$ in our setup. For the YIG sphere 2, its [100] crystal axis is set to be parallel to the bias magnetic field, so we get $K_{2}/2 \pi= 0.0295~{\rm{nHz}}$ and then set $K_{1}$ to be $K_{1}/2 \pi= 0.0074~{\rm{nHz}}$ by adjusting the orientation of the YIG sphere 1.

In the rotating reference frame with respect to the drive-field frequency, we obtain the following quantum Langevin equations:
\begin{equation}\label{sup12}
	\begin{aligned}
		\dot{a}&=-i(\delta_{\rm{c}}-i \kappa_{\rm{c}}) a-i g_{1} b_{1}-g_{2} b_{2}, \\
		\dot{b}_{1}&=-i(\delta_{1}+2K_{1}b_{1}^{\dag}b_{1}+K_{1}-i \gamma_{1}) b_{1}-i g_{1} a+\Omega_{d}, \\
		\dot{b}_{2}&=-i(\delta_{2}+2K_{2}b_{2}^{\dag}b_{2}+K_{2}-i \gamma_{2}) b_{2}+g_{2} a,
	\end{aligned}
\end{equation}
where $\delta_{\rm{c}}=\omega_{\rm{c}}-\omega_{d}$, and $\delta_{1(2)}=\omega_{1(2)}-\omega_{d}$. Then, we write each of the operators $a$, $b_{1}$, and $b_{2}$ as the sum of the steady-state value and the fluctuation, i.e., $a\equiv A+\delta a$, $b_{1}\equiv B_{1}+\delta b_{1}$, and $b_{2}\equiv B_{2}+\delta b_{2}$. At the steady state, we can have the equations for $A$, $B_{1}$, and $B_{2}$:
\begin{equation}\label{sup13}
	\begin{aligned}
		&(\delta_{\rm{c}}-i \kappa_{\rm{c}}) A+g_{1} B_{1}-i g_{2} B_{2}=0, \\
		&(\delta_{1}+\chi_{1}-i \gamma_{1}) B_{1}+g_{1} A+i \Omega_{d}=0, \\
		&(\delta_{2}+\chi_{2}-i \gamma_{2}) B_{2}+i g_{2} A=0,
	\end{aligned}
\end{equation}
where we have used the mean-field approximation for the Kerr terms and the condition $\langle b_{1(2)}^{\dagger} b_{1(2)}\rangle\gg 1$. Here, $\chi_{1(2)}=2 K_{1(2)}\langle b_{1(2)}^{\dagger} b_{1(2)}\rangle$ is the frequency shift of the magnon mode 1(2) due to the Kerr nonlinearity. From the third equation in Eq.~\ref{sup13}, we have
\begin{equation}\label{sup14}
	\begin{aligned}
		B_{2}=-i \eta_{2} (\delta_{2}+\chi_{2}+i \gamma_{2}) A/g_{2},
	\end{aligned}
\end{equation}
where $\eta_{2} ={g_{2}^{2}}/[{(\delta_{2}+\chi_{2})^{2}+\gamma_{2}^{2}}]$. Substituting Eq.~\ref{sup14} into the first equation in Eq.~\ref{sup13}, we have
\begin{equation}\label{sup15}
	\begin{aligned}
		\left[\delta_{c}-\eta_{2}\left(\delta_{2}+\chi_{2}\right)-i\left(\kappa_{c}+\eta_{2} \gamma_{2}\right)\right] \mathrm{A}+g_{1} B_{1}=0.
	\end{aligned}
\end{equation}
Then, we define $\tilde{\delta}_{\rm{c}}\equiv\delta_{\rm{c}}-\eta_{2}(\delta_{2}+\chi_{2})$. Substitution of Eq.~\ref{sup15} into the second equation in Eq.~\ref{sup13} gives
\begin{equation}\label{sup16}
	\begin{aligned}
		\left\{\delta_{1}+\chi_{1}-\eta_{1} \widetilde{\delta}_{c}-i\left[\gamma_{1}+\eta_{1}\left(\kappa_{c}+\eta_{2} \gamma_{2}\right)\right]\right\} B_{1}+i \Omega_{d}=0,
	\end{aligned}
\end{equation}
where $\eta_{1}={g_{1}^{2}}/[{\tilde{\delta}_{\rm{c}}^{2}+(\kappa_{\rm{c}}+\eta_{2} \gamma_{2})^{2}}]$.
Multiplying Eqs.~\ref{sup14}-\ref{sup16} with their complex conjugates, we can obtain the following equations for the frequency shifts of the magnon modes:
\begin{equation}\label{sup17}
	\begin{aligned}
		&[\tilde{\delta}_{\rm{c}}^{2}+(\kappa_{\rm{c}}+\eta_{2} \gamma_{2})^{2}]\Lambda-g_{1}^{2} \chi_{1}=0, \\
		&\{(\delta_{1}+\chi_{1}-\eta_{1} \tilde{\delta}_{\rm{c}})^{2}+[\gamma_{1}+\eta_{1}(\kappa_{\rm{c}}+\eta_{2} \gamma_{2})]^{2}\} \chi_{1}-k P_{d}=0, \\
		&[(\delta_{2}+\chi_{2})^{2}+\gamma_{2}^{2}] \chi_{2}-g_{2}^{2}\varepsilon \Lambda=0,
	\end{aligned}
\end{equation}
where $\Lambda=2K_{1}|A|^{2}$, $\varepsilon= K_{2}/K_{1} $, and $k P_{d}=2K_{1} |\Omega_{d}|^{2}$. This is Eq.~(2) in the main text. It should be noted that $k$ is the only fitting parameter in our experiment. It is related to the conversion efficiency of the drive power $P_{d}$ to the magnon mode 1 via the Kerr effect of magnons. For a given drive power $P_{d}$, we can solve Eq.~\ref{sup17} to obtain the numerical solution for the magnon-mode frequency shift $\chi_{1(2)}$. With these frequency shifts included, we then obtain the effective Hamiltonian of the hybrid system,
\begin{equation}\label{sup18}
	H/\hbar \approx \omega_{\rm{c}} a^{\dagger} a+(\omega_{1}+\chi_{1}) b_{1}^{\dagger} b_{1}+(\omega_{2}+\chi_{2}) b_{2}^{\dagger} b_{2}+g_{1}(a^{\dagger} b_{1}+a b_{1}^{\dagger})-i g_{2}(a^{\dagger} b_{2}-a b_{2}^{\dagger}).
\end{equation}
With given $\chi_{1}$ and $\chi_{2}$, this Hamiltonian has three eigenvalues, i.e., the frequencies of the lower-branch ($\omega_{\rm{LP}}$), middle-branch ($\omega_{\rm{MP}}$), and upper-branch ($\omega_{\rm{UP}}$) cavity magnon polaritons (CMPs). We define the frequency shift of the upper-branch CMP as $\chi_{\rm{UP}}=\omega_{\rm{UP}}(\chi_{1}, \chi_{2})-\omega_{\rm{UP}}(\chi_{1}=0, \chi_{2}=0)$. By numerically solving Eqs.~\ref{sup17} and \ref{sup18}, we can obtain the frequency shift $\chi_{\rm{UP}}$ of the upper-branch CMP. Under appropriate conditions, the frequency shift has five solutions, three of them are stable and other two are unstable.
From the experimental results, we have the following parameters of the hybrid system:
$\omega_{1}/2 \pi=10.212~{\rm{GHz}}$, $\omega_{2}/2 \pi=10.198~{\rm{GHz}}$, $\omega_{\rm{c}}/2 \pi=10.080~{\rm{GHz}}$, $g_{1}/2 \pi=41.5~\rm{MHz}$, $g_{2}/2 \pi=50.5~\rm{MHz}$, $\gamma_{\rm{1}}/2 \pi=13.6~{\rm{MHz}}$, $\gamma_{\rm{2}}/2 \pi=1.3~{\rm{MHz}}$, $\kappa_{\rm{c}}/2 \pi=3.2~{\rm{MHz}}$, $K_{1}/2 \pi= 0.0074~{\rm{nHz}}$, and $K_{2}/2 \pi= 0.0295~{\rm{nHz}}$. Via the multistability experiment, we obtain the parameter $k=20$ by fitting the multistability curves with the numerical results.

\section{\uppercase\expandafter{\romannumeral2}.~The measurement of the multistability}
\begin{figure*}[!htb]
	\renewcommand\thefigure{S\arabic{figure}}
	\centering
	\includegraphics[width=.8\textwidth]{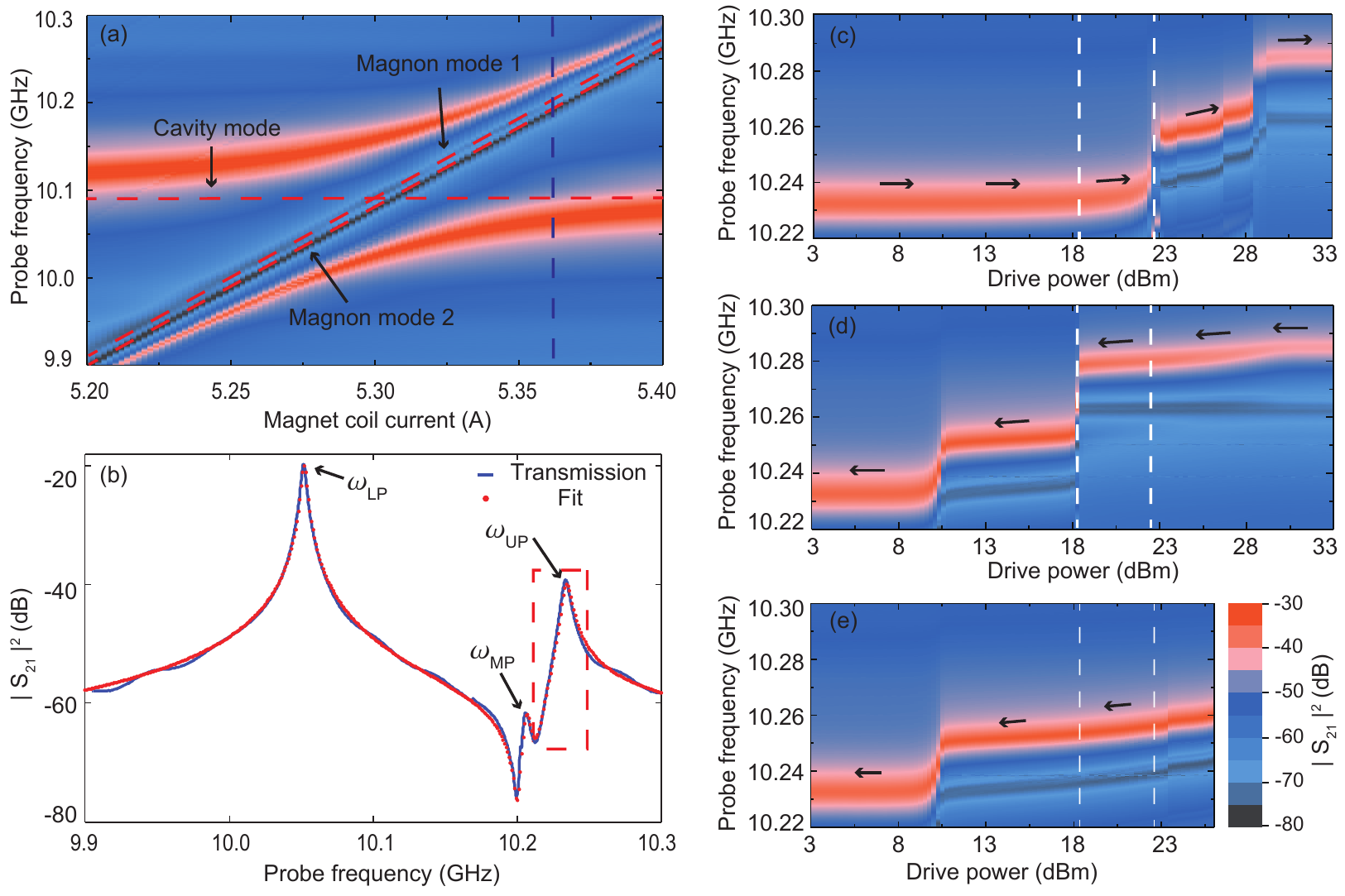}
	\caption{(a) 2D contour plot of the transmission spectra of the cavity magnonic system measured versus both the magnet-coil current (i.e., the bias magnetic field) and the probe field frequency. Two tilted dashed lines indicate the two magnon modes. (b) The transmission spectrum measured at the magnet-coil current 5.37 A. We focus on the upper-branch polariton shown in the red dashed box when demonstrating the multistability in the cavity magnonic system. (c), (d) and (e) The transmission spectra measured versus both the probe field frequency and the drive power while the drive-field frequency detuning is $\delta_{\rm{UP}}/2\pi=-18~{\rm{MHz}}$. The red peak corresponds to the upper-branch CMP. (c) The drive power is swept forward. (d) The drive power is swept backward. (e) The drive power is swept backward from 26 dBm, and the upper-branch CMP is initially prepared in the frequency-shift middle stable state (MSS). The tristable region can be found between the two vertical dashed lines. The color scale is the same for all graphs.}
	\label{Fig:S1}
\end{figure*}

To get the coupling strengths, in the experiment, we first place YIG sphere 1 in the cavity and measure the transmission spectrum. The coupling strength between the Kittel mode in YIG sphere 1 and the cavity mode is determined at the resonance point, which is $g_{1}/2 \pi=41.5~\rm{MHz}$. Then we place YIG sphere 2 in the cavity without removing YIG sphere 1 out. The coupling strength between the Kittel mode in YIG sphere 2 and the cavity mode is obtained as $g_{2}/2 \pi=50.5~\rm{MHz}$ by fitting the transmission curve shown in Fig.~\ref{Fig:S1}(b).

The transmission spectra of the cavity magnonic system without applying a drive field are measured and plotted versus both the magnet-coil current and the probe-field frequency, as shown in Fig.~\ref{Fig:S1}(a), where three branches of CMP occur. When the magnet-coil current is 5.37~A, the frequencies of the two magnon modes are $\omega_{1}/2 \pi=10.212~$GHz and $\omega_{2}/2 \pi=10.198~$GHz. In this case, the three branches of CMP are labeled in Fig.~\ref{Fig:S1}(b). The frequencies of the lower-branch, middle-branch, and upper-branch CMPs are denoted as $\omega_{\rm{LP}}$, $\omega_{\rm{MP}}$, and  $\omega_{\rm{UP}}$, respectively.
We apply the drive field on the upper-branch CMP with $\delta_{\rm{UP}}/ 2 \pi=(\omega_{\rm{UP}} -\omega_{d})/ 2 \pi=-18$ ~MHz and measure the transmission spectra by sweeping the drive-field power. The measured 2D contour plots are shown in Figs.~\ref{Fig:S1}(c)-\ref{Fig:S1}(e), where the peaks in red represent the upper-branch CMP. The central frequencies of the peaks correspond to the frequency of the upper-branch polariton. We use the `max' function in Matlab$^\circledR$ to find the peak frequencies of the transmission spectra. Figures~\ref{Fig:S1}(c) and~\ref{Fig:S1}(d) are obtained by sweeping the drive power forward and backward, respectively. The frequency shifts and jumps of the upper-branch CMP are clearly observed. Figure~\ref{Fig:S1}(e) shows the result obtained by sweeping the drive power backward when the upper-branch CMP is initially prepared in the middle stable state of the frequency shift. Figures~\ref{Fig:S1}(c)-\ref{Fig:S1}(e) clearly indicate that there are three stable states for the frequency shift of the upper-branch CMP when the drive power ranges from 18.5 to 22 dBm. Therefore, we have experimentally realized the multistability of the CMP.

\section{\uppercase\expandafter{\romannumeral3}. The frequency-shift relaxation rate}

Following Eq.~\ref{sup4}, the Hamiltonian for the magnetocrystalline anisotropy energy of the YIG sphere is
\begin{equation}\label{sup19}
	H_{\rm{an}}/\hbar=-\frac{\hbar K_{\rm{an}}\gamma^2}{M^{2}V_{\rm{m}}}S_{z}^2.
\end{equation}
Using the Holstein-Primakoff transformation~\cite{Holstein40} $S_{z}=S-b^{\dag}b$, we have
\begin{equation}\label{sup20}
	H_{\rm{an}}/\hbar=-\frac{\hbar K_{\rm{an}}\gamma^2}{M^{2}V_{\rm{m}}} (S^2-2Sb^{\dagger} b +b^{\dagger} b b^{\dagger} b).
\end{equation}
In the low-excitation case with $\langle b^{\dag}b\rangle\ll S$, $H_{\rm{an}}$ can be approximated as
\begin{equation}\label{sup21}
	H_{\rm{an}}/\hbar \approx -\frac{\hbar K_{\rm{an}}\gamma^2}{M^{2}V_{\rm{m}}} (S^2-2Sb^{\dagger} b).
\end{equation}
Without the magnon excitation, the magnetocrystalline anisotropy energy is $H_{\rm{an}}/\hbar=-{\hbar K_{\rm{an}}\gamma^2S^2}/({M^{2}V_{\rm{m}}})$. Under the mean-field approximation, the variation of the magnetocrystalline anisotropy energy due to the magnon excitation can be written as
\begin{equation}\label{sup22}
	\Delta E_{\rm{an}}/\hbar\approx\frac{2S\hbar K_{\rm{an}}\gamma^2}{M^{2}V_{\rm{m}}} \langle b^{\dagger} b\rangle=2SK\langle b^{\dagger} b\rangle,
\end{equation}
which is proportional to the magnon excitation number.

\begin{figure*}[!htp]
	\renewcommand\thefigure{S\arabic{figure}}
	\centering
	\includegraphics{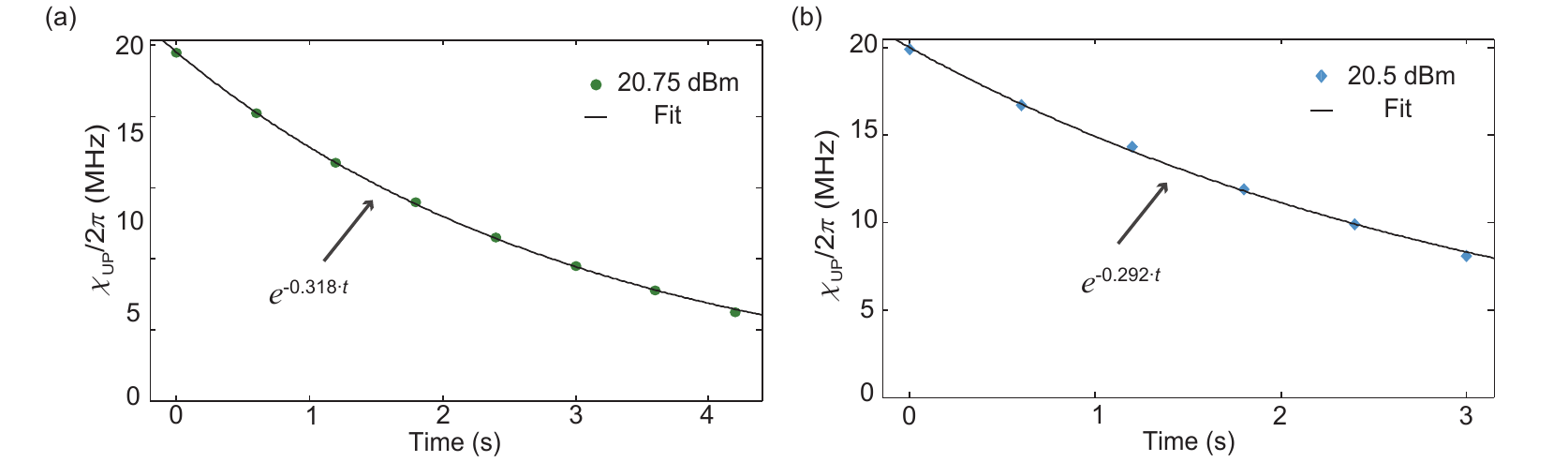}
	\caption{Relaxation rate measurements for the CMP frequency shift. (a) The frequency shift of the upper-branch CMP is first tuned to one MSS where the drive power is 20.75~dBm. The green circles are the experimental results, and the black line is the exponential fitting curve. The relaxation rate is fitted to be $\Gamma_{\rm{UP}}/2\pi=0.051~\rm{Hz}$. (b) The initial state is another MSS where the drive power is 20.5~dBm. The blue rhombuses are the experimental results. The relaxation rate is fitted to be $\Gamma_{\rm{UP}}/2\pi=0.046~\rm{Hz}$.}
	\label{Fig:S2}
\end{figure*}

In the experiment, when the drive field is turned off, the magnon excitation number will decrease and the magnetocrystalline anisotropy energy will relax. Since the frequency shift of the magnon mode is also proportional to the magnon excitation number, the relaxation rate of the frequency shift reflects the time scale of the magnetocrystalline anisotropy energy relaxation. In the main text, we have shown that when the upper-branch CMP is driven to the middle stable state (MSS) where the drive power is 21 dBm, the measured energy relaxation rate is $0.31~$Hz. In Fig.~\ref{Fig:S2}, we measure the magnetocrystalline anisotropy energy relaxation after the upper-branch CMP is prepared in the MSS by using two different drive powers. For the MSS with the drive power of 20.75 dBm, the energy relaxation rate is $0.051$~Hz. For the MSS with the drive power of 20.5 dBm, the energy relaxation rate is $0.046$~Hz. As expected, the measured energy relaxation rates are very close to each other. These individual measurements reveal the intrinsic relaxation rate of the magnetocrystalline anisotropy energy.

\begin{figure*}[!h]
	\renewcommand\thefigure{S\arabic{figure}}
	\centering
	\includegraphics{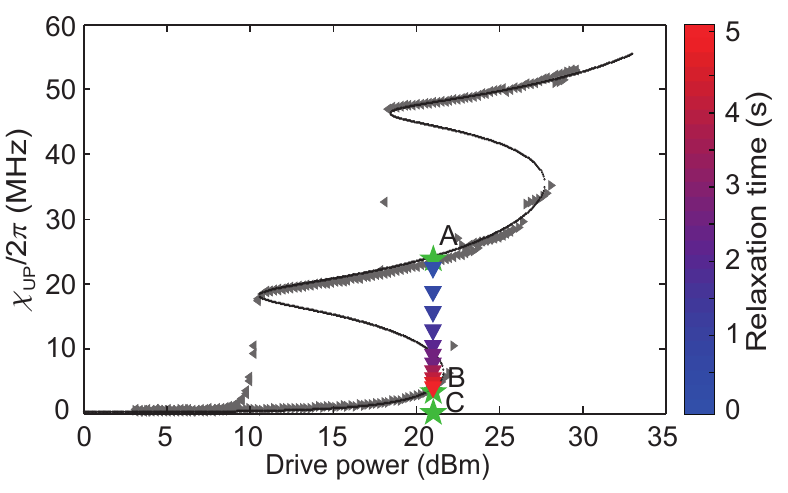}
	\caption{Relaxation of the upper-branch CMP frequency shift in the tristable region. The critical memory time is determined by the relaxation time of the frequency shift from A to B.}
	\label{Fig:S3}
\end{figure*}

\section{\uppercase\expandafter{\romannumeral4}. Relationship between the maximum memory time and the relaxation time of the CMP frequency shift}

As shown in Fig.~\ref{Fig:S3}, the gray triangles are the frequency shift $\chi_{\rm UP}$ of the upper-branch CMP at different drive powers. The data points are the same as in Fig. 2(d) in the main text. After the frequency shift is prepared to be at the point A, we turn off the drive field. The frequency shift first relaxes to B, and then to C (zero frequency shift) afterwards. When we turn on the drive field before the frequency shift goes down below B, the frequency shift can still go back to A. Therefore, the state A is revived. When we turn on the drive field after the frequency shift goes down below B, the frequency shift can only go back to B even if we turn on the drive field again. Then, the memory of the state A is lost.

The maximum memory time is determined by the relaxation time of the CMP frequency shift from A to B, as shown in Fig.~\ref{Fig:S3}, which is 5.11~s in the present case.


\begin{thebibliography}{59}%
	\makeatletter
	\providecommand \@ifxundefined [1]{%
		\@ifx{#1\undefined}
	}%
	\providecommand \@ifnum [1]{%
		\ifnum #1\expandafter \@firstoftwo
		\else \expandafter \@secondoftwo
		\fi
	}%
	\providecommand \@ifx [1]{%
		\ifx #1\expandafter \@firstoftwo
		\else \expandafter \@secondoftwo
		\fi
	}%
	\providecommand \natexlab [1]{#1}%
	\providecommand \enquote  [1]{``#1''}%
	\providecommand \bibnamefont  [1]{#1}%
	\providecommand \bibfnamefont [1]{#1}%
	\providecommand \citenamefont [1]{#1}%
	\providecommand \href@noop [0]{\@secondoftwo}%
	\providecommand \href [0]{\begingroup \@sanitize@url \@href}%
	\providecommand \@href[1]{\@@startlink{#1}\@@href}%
	\providecommand \@@href[1]{\endgroup#1\@@endlink}%
	\providecommand \@sanitize@url [0]{\catcode `\\12\catcode `\$12\catcode
		`\&12\catcode `\#12\catcode `\^12\catcode `\_12\catcode `\%12\relax}%
	\providecommand \@@startlink[1]{}%
	\providecommand \@@endlink[0]{}%
	\providecommand \url  [0]{\begingroup\@sanitize@url \@url }%
	\providecommand \@url [1]{\endgroup\@href {#1}{\urlprefix }}%
	\providecommand \urlprefix  [0]{URL }%
	\providecommand \Eprint [0]{\href }%
	\providecommand \doibase [0]{http://dx.doi.org/}%
	\providecommand \selectlanguage [0]{\@gobble}%
	\providecommand \bibinfo  [0]{\@secondoftwo}%
	\providecommand \bibfield  [0]{\@secondoftwo}%
	\providecommand \translation [1]{[#1]}%
	\providecommand \BibitemOpen [0]{}%
	\providecommand \bibitemStop [0]{}%
	\providecommand \bibitemNoStop [0]{.\EOS\space}%
	\providecommand \EOS [0]{\spacefactor3000\relax}%
	\providecommand \BibitemShut  [1]{\csname bibitem#1\endcsname}%
	\let\auto@bib@innerbib\@empty
	%</preamble>
	\bibitem [{\citenamefont {Forn-D\'{\i}az}\ \emph {et~al.}(2019)\citenamefont
		{Forn-D\'{\i}az}, \citenamefont {Lamata}, \citenamefont {Rico}, \citenamefont
		{Kono},\ and\ \citenamefont {Solano}}]{RevModPhys.91.025005}%
	\BibitemOpen
	\bibfield  {author} {\bibinfo {author} {\bibfnamefont {P.}~\bibnamefont
			{Forn-D\'{\i}az}}, \bibinfo {author} {\bibfnamefont {L.}~\bibnamefont
			{Lamata}}, \bibinfo {author} {\bibfnamefont {E.}~\bibnamefont {Rico}},
		\bibinfo {author} {\bibfnamefont {J.}~\bibnamefont {Kono}}, \ and\ \bibinfo
		{author} {\bibfnamefont {E.}~\bibnamefont {Solano}},\ }\href {\doibase
		10.1103/RevModPhys.91.025005} {\bibfield  {journal} {\bibinfo  {journal}
			{Rev. Mod. Phys.}\ }\textbf {\bibinfo {volume} {91}},\ \bibinfo {pages}
		{025005} (\bibinfo {year} {2019})}\BibitemShut {NoStop}%
	\bibitem [{\citenamefont {Frisk~Kockum}\ \emph {et~al.}(2019)\citenamefont
		{Frisk~Kockum}, \citenamefont {Miranowicz}, \citenamefont {De~Liberato},
		\citenamefont {Savasta},\ and\ \citenamefont {Nori}}]{FriskKockum2019}%
	\BibitemOpen
	\bibfield  {author} {\bibinfo {author} {\bibfnamefont {A.}~\bibnamefont
			{Frisk~Kockum}}, \bibinfo {author} {\bibfnamefont {A.}~\bibnamefont
			{Miranowicz}}, \bibinfo {author} {\bibfnamefont {S.}~\bibnamefont
			{De~Liberato}}, \bibinfo {author} {\bibfnamefont {S.}~\bibnamefont
			{Savasta}}, \ and\ \bibinfo {author} {\bibfnamefont {F.}~\bibnamefont
			{Nori}},\ }\href {\doibase 10.1038/s42254-018-0006-2} {\bibfield  {journal}
		{\bibinfo  {journal} {Nature Reviews Physics}\ }\textbf {\bibinfo {volume}
			{1}},\ \bibinfo {pages} {19} (\bibinfo {year} {2019})}\BibitemShut {NoStop}%
	\bibitem [{\citenamefont {Yu}\ \emph {et~al.}(2019{\natexlab{a}})\citenamefont
		{Yu}, \citenamefont {Peng}, \citenamefont {Yang},\ and\ \citenamefont
		{Li}}]{Yu2019}%
	\BibitemOpen
	\bibfield  {author} {\bibinfo {author} {\bibfnamefont {H.}~\bibnamefont
			{Yu}}, \bibinfo {author} {\bibfnamefont {Y.}~\bibnamefont {Peng}}, \bibinfo
		{author} {\bibfnamefont {Y.}~\bibnamefont {Yang}}, \ and\ \bibinfo {author}
		{\bibfnamefont {Z.-Y.}\ \bibnamefont {Li}},\ }\href {\doibase
		10.1038/s41524-019-0184-1} {\bibfield  {journal} {\bibinfo  {journal} {npj
				Computational Materials}\ }\textbf {\bibinfo {volume} {5}},\ \bibinfo {pages}
		{45} (\bibinfo {year} {2019}{\natexlab{a}})}\BibitemShut {NoStop}%
	\bibitem [{\citenamefont {Ayuso}\ \emph {et~al.}(2019)\citenamefont {Ayuso},
		\citenamefont {Neufeld}, \citenamefont {Ordonez}, \citenamefont {Decleva},
		\citenamefont {Lerner}, \citenamefont {Cohen}, \citenamefont {Ivanov},\ and\
		\citenamefont {Smirnova}}]{Ayuso2019}%
	\BibitemOpen
	\bibfield  {author} {\bibinfo {author} {\bibfnamefont {D.}~\bibnamefont
			{Ayuso}}, \bibinfo {author} {\bibfnamefont {O.}~\bibnamefont {Neufeld}},
		\bibinfo {author} {\bibfnamefont {A.~F.}\ \bibnamefont {Ordonez}}, \bibinfo
		{author} {\bibfnamefont {P.}~\bibnamefont {Decleva}}, \bibinfo {author}
		{\bibfnamefont {G.}~\bibnamefont {Lerner}}, \bibinfo {author} {\bibfnamefont
			{O.}~\bibnamefont {Cohen}}, \bibinfo {author} {\bibfnamefont
			{M.}~\bibnamefont {Ivanov}}, \ and\ \bibinfo {author} {\bibfnamefont
			{O.}~\bibnamefont {Smirnova}},\ }\href {\doibase 10.1038/s41566-019-0531-2}
	{\bibfield  {journal} {\bibinfo  {journal} {Nature Photonics}\ }\textbf
		{\bibinfo {volume} {13}},\ \bibinfo {pages} {866} (\bibinfo {year}
		{2019})}\BibitemShut {NoStop}%
	\bibitem [{\citenamefont {Rivera}\ and\ \citenamefont
		{Kaminer}(2020)}]{Rivera2020}%
	\BibitemOpen
	\bibfield  {author} {\bibinfo {author} {\bibfnamefont {N.}~\bibnamefont
			{Rivera}}\ and\ \bibinfo {author} {\bibfnamefont {I.}~\bibnamefont
			{Kaminer}},\ }\href {\doibase 10.1038/s42254-020-0224-2} {\bibfield
		{journal} {\bibinfo  {journal} {Nature Reviews Physics}\ }\textbf {\bibinfo
			{volume} {2}},\ \bibinfo {pages} {538} (\bibinfo {year} {2020})}\BibitemShut
	{NoStop}%
	\bibitem [{\citenamefont {Huebl}\ \emph {et~al.}(2013)\citenamefont {Huebl},
		\citenamefont {Zollitsch}, \citenamefont {Lotze}, \citenamefont {Hocke},
		\citenamefont {Greifenstein}, \citenamefont {Marx}, \citenamefont {Gross},\
		and\ \citenamefont {Goennenwein}}]{PhysRevLett.111.127003}%
	\BibitemOpen
	\bibfield  {author} {\bibinfo {author} {\bibfnamefont {H.}~\bibnamefont
			{Huebl}}, \bibinfo {author} {\bibfnamefont {C.~W.}\ \bibnamefont
			{Zollitsch}}, \bibinfo {author} {\bibfnamefont {J.}~\bibnamefont {Lotze}},
		\bibinfo {author} {\bibfnamefont {F.}~\bibnamefont {Hocke}}, \bibinfo
		{author} {\bibfnamefont {M.}~\bibnamefont {Greifenstein}}, \bibinfo {author}
		{\bibfnamefont {A.}~\bibnamefont {Marx}}, \bibinfo {author} {\bibfnamefont
			{R.}~\bibnamefont {Gross}}, \ and\ \bibinfo {author} {\bibfnamefont
			{S.~T.~B.}\ \bibnamefont {Goennenwein}},\ }\href {\doibase
		10.1103/PhysRevLett.111.127003} {\bibfield  {journal} {\bibinfo  {journal}
			{Phys. Rev. Lett.}\ }\textbf {\bibinfo {volume} {111}},\ \bibinfo {pages}
		{127003} (\bibinfo {year} {2013})}\BibitemShut {NoStop}%
	\bibitem [{\citenamefont {Tabuchi}\ \emph {et~al.}(2014)\citenamefont
		{Tabuchi}, \citenamefont {Ishino}, \citenamefont {Ishikawa}, \citenamefont
		{Yamazaki}, \citenamefont {Usami},\ and\ \citenamefont
		{Nakamura}}]{PhysRevLett.113.083603}%
	\BibitemOpen
	\bibfield  {author} {\bibinfo {author} {\bibfnamefont {Y.}~\bibnamefont
			{Tabuchi}}, \bibinfo {author} {\bibfnamefont {S.}~\bibnamefont {Ishino}},
		\bibinfo {author} {\bibfnamefont {T.}~\bibnamefont {Ishikawa}}, \bibinfo
		{author} {\bibfnamefont {R.}~\bibnamefont {Yamazaki}}, \bibinfo {author}
		{\bibfnamefont {K.}~\bibnamefont {Usami}}, \ and\ \bibinfo {author}
		{\bibfnamefont {Y.}~\bibnamefont {Nakamura}},\ }\href {\doibase
		10.1103/PhysRevLett.113.083603} {\bibfield  {journal} {\bibinfo  {journal}
			{Phys. Rev. Lett.}\ }\textbf {\bibinfo {volume} {113}},\ \bibinfo {pages}
		{083603} (\bibinfo {year} {2014})}\BibitemShut {NoStop}%
	\bibitem [{\citenamefont {Zhang}\ \emph {et~al.}(2014)\citenamefont {Zhang},
		\citenamefont {Zou}, \citenamefont {Jiang},\ and\ \citenamefont
		{Tang}}]{PhysRevLett.113.156401}%
	\BibitemOpen
	\bibfield  {author} {\bibinfo {author} {\bibfnamefont {X.}~\bibnamefont
			{Zhang}}, \bibinfo {author} {\bibfnamefont {C.-L.}\ \bibnamefont {Zou}},
		\bibinfo {author} {\bibfnamefont {L.}~\bibnamefont {Jiang}}, \ and\ \bibinfo
		{author} {\bibfnamefont {H.~X.}\ \bibnamefont {Tang}},\ }\href {\doibase
		10.1103/PhysRevLett.113.156401} {\bibfield  {journal} {\bibinfo  {journal}
			{Phys. Rev. Lett.}\ }\textbf {\bibinfo {volume} {113}},\ \bibinfo {pages}
		{156401} (\bibinfo {year} {2014})}\BibitemShut {NoStop}%
	\bibitem [{\citenamefont {Goryachev}\ \emph {et~al.}(2014)\citenamefont
		{Goryachev}, \citenamefont {Farr}, \citenamefont {Creedon}, \citenamefont
		{Fan}, \citenamefont {Kostylev},\ and\ \citenamefont
		{Tobar}}]{PhysRevApplied.2.054002}%
	\BibitemOpen
	\bibfield  {author} {\bibinfo {author} {\bibfnamefont {M.}~\bibnamefont
			{Goryachev}}, \bibinfo {author} {\bibfnamefont {W.~G.}\ \bibnamefont {Farr}},
		\bibinfo {author} {\bibfnamefont {D.~L.}\ \bibnamefont {Creedon}}, \bibinfo
		{author} {\bibfnamefont {Y.}~\bibnamefont {Fan}}, \bibinfo {author}
		{\bibfnamefont {M.}~\bibnamefont {Kostylev}}, \ and\ \bibinfo {author}
		{\bibfnamefont {M.~E.}\ \bibnamefont {Tobar}},\ }\href {\doibase
		10.1103/PhysRevApplied.2.054002} {\bibfield  {journal} {\bibinfo  {journal}
			{Phys. Rev. Applied}\ }\textbf {\bibinfo {volume} {2}},\ \bibinfo {pages}
		{054002} (\bibinfo {year} {2014})}\BibitemShut {NoStop}%
	\bibitem [{\citenamefont {Bai}\ \emph {et~al.}(2015)\citenamefont {Bai},
		\citenamefont {Harder}, \citenamefont {Chen}, \citenamefont {Fan},
		\citenamefont {Xiao},\ and\ \citenamefont {Hu}}]{PhysRevLett.114.227201}%
	\BibitemOpen
	\bibfield  {author} {\bibinfo {author} {\bibfnamefont {L.}~\bibnamefont
			{Bai}}, \bibinfo {author} {\bibfnamefont {M.}~\bibnamefont {Harder}},
		\bibinfo {author} {\bibfnamefont {Y.~P.}\ \bibnamefont {Chen}}, \bibinfo
		{author} {\bibfnamefont {X.}~\bibnamefont {Fan}}, \bibinfo {author}
		{\bibfnamefont {J.~Q.}\ \bibnamefont {Xiao}}, \ and\ \bibinfo {author}
		{\bibfnamefont {C.-M.}\ \bibnamefont {Hu}},\ }\href {\doibase
		10.1103/PhysRevLett.114.227201} {\bibfield  {journal} {\bibinfo  {journal}
			{Phys. Rev. Lett.}\ }\textbf {\bibinfo {volume} {114}},\ \bibinfo {pages}
		{227201} (\bibinfo {year} {2015})}\BibitemShut {NoStop}%
	\bibitem [{\citenamefont {Cao}\ \emph {et~al.}(2015)\citenamefont {Cao},
		\citenamefont {Yan}, \citenamefont {Huebl}, \citenamefont {Goennenwein},\
		and\ \citenamefont {Bauer}}]{PhysRevB.91.094423}%
	\BibitemOpen
	\bibfield  {author} {\bibinfo {author} {\bibfnamefont {Y.}~\bibnamefont
			{Cao}}, \bibinfo {author} {\bibfnamefont {P.}~\bibnamefont {Yan}}, \bibinfo
		{author} {\bibfnamefont {H.}~\bibnamefont {Huebl}}, \bibinfo {author}
		{\bibfnamefont {S.~T.~B.}\ \bibnamefont {Goennenwein}}, \ and\ \bibinfo
		{author} {\bibfnamefont {G.~E.~W.}\ \bibnamefont {Bauer}},\ }\href {\doibase
		10.1103/PhysRevB.91.094423} {\bibfield  {journal} {\bibinfo  {journal} {Phys.
				Rev. B}\ }\textbf {\bibinfo {volume} {91}},\ \bibinfo {pages} {094423}
		(\bibinfo {year} {2015})}\BibitemShut {NoStop}%
	\bibitem [{\citenamefont {Zhang}\ \emph {et~al.}(2015)\citenamefont {Zhang},
		\citenamefont {Zou}, \citenamefont {Zhu}, \citenamefont {Marquardt},
		\citenamefont {Jiang},\ and\ \citenamefont {Tang}}]{Zhang2015}%
	\BibitemOpen
	\bibfield  {author} {\bibinfo {author} {\bibfnamefont {X.}~\bibnamefont
			{Zhang}}, \bibinfo {author} {\bibfnamefont {C.-L.}\ \bibnamefont {Zou}},
		\bibinfo {author} {\bibfnamefont {N.}~\bibnamefont {Zhu}}, \bibinfo {author}
		{\bibfnamefont {F.}~\bibnamefont {Marquardt}}, \bibinfo {author}
		{\bibfnamefont {L.}~\bibnamefont {Jiang}}, \ and\ \bibinfo {author}
		{\bibfnamefont {H.~X.}\ \bibnamefont {Tang}},\ }\href {\doibase
		10.1038/ncomms9914} {\bibfield  {journal} {\bibinfo  {journal} {Nature
				Communications}\ }\textbf {\bibinfo {volume} {6}},\ \bibinfo {pages} {8914}
		(\bibinfo {year} {2015})}\BibitemShut {NoStop}%
	\bibitem [{\citenamefont {Tabuchi}\ \emph {et~al.}(2015)\citenamefont
		{Tabuchi}, \citenamefont {Ishino}, \citenamefont {Noguchi}, \citenamefont
		{Ishikawa}, \citenamefont {Yamazaki}, \citenamefont {Usami},\ and\
		\citenamefont {Nakamura}}]{Tabuchi405}%
	\BibitemOpen
	\bibfield  {author} {\bibinfo {author} {\bibfnamefont {Y.}~\bibnamefont
			{Tabuchi}}, \bibinfo {author} {\bibfnamefont {S.}~\bibnamefont {Ishino}},
		\bibinfo {author} {\bibfnamefont {A.}~\bibnamefont {Noguchi}}, \bibinfo
		{author} {\bibfnamefont {T.}~\bibnamefont {Ishikawa}}, \bibinfo {author}
		{\bibfnamefont {R.}~\bibnamefont {Yamazaki}}, \bibinfo {author}
		{\bibfnamefont {K.}~\bibnamefont {Usami}}, \ and\ \bibinfo {author}
		{\bibfnamefont {Y.}~\bibnamefont {Nakamura}},\ }\href {\doibase
		10.1126/science.aaa3693} {\bibfield  {journal} {\bibinfo  {journal}
			{Science}\ }\textbf {\bibinfo {volume} {349}},\ \bibinfo {pages} {405}
		(\bibinfo {year} {2015})}\BibitemShut {NoStop}%
	\bibitem [{\citenamefont {Haigh}\ \emph {et~al.}(2016)\citenamefont {Haigh},
		\citenamefont {Nunnenkamp}, \citenamefont {Ramsay},\ and\ \citenamefont
		{Ferguson}}]{PhysRevLett.117.133602}%
	\BibitemOpen
	\bibfield  {author} {\bibinfo {author} {\bibfnamefont {J.~A.}\ \bibnamefont
			{Haigh}}, \bibinfo {author} {\bibfnamefont {A.}~\bibnamefont {Nunnenkamp}},
		\bibinfo {author} {\bibfnamefont {A.~J.}\ \bibnamefont {Ramsay}}, \ and\
		\bibinfo {author} {\bibfnamefont {A.~J.}\ \bibnamefont {Ferguson}},\ }\href
	{\doibase 10.1103/PhysRevLett.117.133602} {\bibfield  {journal} {\bibinfo
			{journal} {Phys. Rev. Lett.}\ }\textbf {\bibinfo {volume} {117}},\ \bibinfo
		{pages} {133602} (\bibinfo {year} {2016})}\BibitemShut {NoStop}%
	\bibitem [{\citenamefont {Maier-Flaig}\ \emph {et~al.}(2016)\citenamefont
		{Maier-Flaig}, \citenamefont {Harder}, \citenamefont {Gross}, \citenamefont
		{Huebl},\ and\ \citenamefont {Goennenwein}}]{PhysRevB.94.054433}%
	\BibitemOpen
	\bibfield  {author} {\bibinfo {author} {\bibfnamefont {H.}~\bibnamefont
			{Maier-Flaig}}, \bibinfo {author} {\bibfnamefont {M.}~\bibnamefont {Harder}},
		\bibinfo {author} {\bibfnamefont {R.}~\bibnamefont {Gross}}, \bibinfo
		{author} {\bibfnamefont {H.}~\bibnamefont {Huebl}}, \ and\ \bibinfo {author}
		{\bibfnamefont {S.~T.~B.}\ \bibnamefont {Goennenwein}},\ }\href {\doibase
		10.1103/PhysRevB.94.054433} {\bibfield  {journal} {\bibinfo  {journal} {Phys.
				Rev. B}\ }\textbf {\bibinfo {volume} {94}},\ \bibinfo {pages} {054433}
		(\bibinfo {year} {2016})}\BibitemShut {NoStop}%
	\bibitem [{\citenamefont {Bourhill}\ \emph {et~al.}(2016)\citenamefont
		{Bourhill}, \citenamefont {Kostylev}, \citenamefont {Goryachev},
		\citenamefont {Creedon},\ and\ \citenamefont {Tobar}}]{PhysRevB.93.144420}%
	\BibitemOpen
	\bibfield  {author} {\bibinfo {author} {\bibfnamefont {J.}~\bibnamefont
			{Bourhill}}, \bibinfo {author} {\bibfnamefont {N.}~\bibnamefont {Kostylev}},
		\bibinfo {author} {\bibfnamefont {M.}~\bibnamefont {Goryachev}}, \bibinfo
		{author} {\bibfnamefont {D.~L.}\ \bibnamefont {Creedon}}, \ and\ \bibinfo
		{author} {\bibfnamefont {M.~E.}\ \bibnamefont {Tobar}},\ }\href {\doibase
		10.1103/PhysRevB.93.144420} {\bibfield  {journal} {\bibinfo  {journal} {Phys.
				Rev. B}\ }\textbf {\bibinfo {volume} {93}},\ \bibinfo {pages} {144420}
		(\bibinfo {year} {2016})}\BibitemShut {NoStop}%
	\bibitem [{\citenamefont {Hisatomi}\ \emph {et~al.}(2016)\citenamefont
		{Hisatomi}, \citenamefont {Osada}, \citenamefont {Tabuchi}, \citenamefont
		{Ishikawa}, \citenamefont {Noguchi}, \citenamefont {Yamazaki}, \citenamefont
		{Usami},\ and\ \citenamefont {Nakamura}}]{PhysRevB.93.174427}%
	\BibitemOpen
	\bibfield  {author} {\bibinfo {author} {\bibfnamefont {R.}~\bibnamefont
			{Hisatomi}}, \bibinfo {author} {\bibfnamefont {A.}~\bibnamefont {Osada}},
		\bibinfo {author} {\bibfnamefont {Y.}~\bibnamefont {Tabuchi}}, \bibinfo
		{author} {\bibfnamefont {T.}~\bibnamefont {Ishikawa}}, \bibinfo {author}
		{\bibfnamefont {A.}~\bibnamefont {Noguchi}}, \bibinfo {author} {\bibfnamefont
			{R.}~\bibnamefont {Yamazaki}}, \bibinfo {author} {\bibfnamefont
			{K.}~\bibnamefont {Usami}}, \ and\ \bibinfo {author} {\bibfnamefont
			{Y.}~\bibnamefont {Nakamura}},\ }\href {\doibase 10.1103/PhysRevB.93.174427}
	{\bibfield  {journal} {\bibinfo  {journal} {Phys. Rev. B}\ }\textbf {\bibinfo
			{volume} {93}},\ \bibinfo {pages} {174427} (\bibinfo {year}
		{2016})}\BibitemShut {NoStop}%
	\bibitem [{\citenamefont {Osada}\ \emph {et~al.}(2016)\citenamefont {Osada},
		\citenamefont {Hisatomi}, \citenamefont {Noguchi}, \citenamefont {Tabuchi},
		\citenamefont {Yamazaki}, \citenamefont {Usami}, \citenamefont {Sadgrove},
		\citenamefont {Yalla}, \citenamefont {Nomura},\ and\ \citenamefont
		{Nakamura}}]{PhysRevLett.116.223601}%
	\BibitemOpen
	\bibfield  {author} {\bibinfo {author} {\bibfnamefont {A.}~\bibnamefont
			{Osada}}, \bibinfo {author} {\bibfnamefont {R.}~\bibnamefont {Hisatomi}},
		\bibinfo {author} {\bibfnamefont {A.}~\bibnamefont {Noguchi}}, \bibinfo
		{author} {\bibfnamefont {Y.}~\bibnamefont {Tabuchi}}, \bibinfo {author}
		{\bibfnamefont {R.}~\bibnamefont {Yamazaki}}, \bibinfo {author}
		{\bibfnamefont {K.}~\bibnamefont {Usami}}, \bibinfo {author} {\bibfnamefont
			{M.}~\bibnamefont {Sadgrove}}, \bibinfo {author} {\bibfnamefont
			{R.}~\bibnamefont {Yalla}}, \bibinfo {author} {\bibfnamefont
			{M.}~\bibnamefont {Nomura}}, \ and\ \bibinfo {author} {\bibfnamefont
			{Y.}~\bibnamefont {Nakamura}},\ }\href {\doibase
		10.1103/PhysRevLett.116.223601} {\bibfield  {journal} {\bibinfo  {journal}
			{Phys. Rev. Lett.}\ }\textbf {\bibinfo {volume} {116}},\ \bibinfo {pages}
		{223601} (\bibinfo {year} {2016})}\BibitemShut {NoStop}%
	\bibitem [{\citenamefont {Zhang}\ \emph {et~al.}(2016)\citenamefont {Zhang},
		\citenamefont {Zhu}, \citenamefont {Zou},\ and\ \citenamefont
		{Tang}}]{PhysRevLett.117.123605}%
	\BibitemOpen
	\bibfield  {author} {\bibinfo {author} {\bibfnamefont {X.}~\bibnamefont
			{Zhang}}, \bibinfo {author} {\bibfnamefont {N.}~\bibnamefont {Zhu}}, \bibinfo
		{author} {\bibfnamefont {C.-L.}\ \bibnamefont {Zou}}, \ and\ \bibinfo
		{author} {\bibfnamefont {H.~X.}\ \bibnamefont {Tang}},\ }\href {\doibase
		10.1103/PhysRevLett.117.123605} {\bibfield  {journal} {\bibinfo  {journal}
			{Phys. Rev. Lett.}\ }\textbf {\bibinfo {volume} {117}},\ \bibinfo {pages}
		{123605} (\bibinfo {year} {2016})}\BibitemShut {NoStop}%
	\bibitem [{\citenamefont {Zhang}\ \emph {et~al.}(2017)\citenamefont {Zhang},
		\citenamefont {Luo}, \citenamefont {Wang}, \citenamefont {Li},\ and\
		\citenamefont {You}}]{Zhang2017}%
	\BibitemOpen
	\bibfield  {author} {\bibinfo {author} {\bibfnamefont {D.}~\bibnamefont
			{Zhang}}, \bibinfo {author} {\bibfnamefont {X.-Q.}\ \bibnamefont {Luo}},
		\bibinfo {author} {\bibfnamefont {Y.-P.}\ \bibnamefont {Wang}}, \bibinfo
		{author} {\bibfnamefont {T.-F.}\ \bibnamefont {Li}}, \ and\ \bibinfo {author}
		{\bibfnamefont {J.~Q.}\ \bibnamefont {You}},\ }\href {\doibase
		10.1038/s41467-017-01634-w} {\bibfield  {journal} {\bibinfo  {journal}
			{Nature Communications}\ }\textbf {\bibinfo {volume} {8}},\ \bibinfo {pages}
		{1368} (\bibinfo {year} {2017})}\BibitemShut {NoStop}%
	\bibitem [{\citenamefont {Wang}\ \emph {et~al.}(2018)\citenamefont {Wang},
		\citenamefont {Zhang}, \citenamefont {Zhang}, \citenamefont {Li},
		\citenamefont {Hu},\ and\ \citenamefont {You}}]{PhysRevLett.120.057202}%
	\BibitemOpen
	\bibfield  {author} {\bibinfo {author} {\bibfnamefont {Y.-P.}\ \bibnamefont
			{Wang}}, \bibinfo {author} {\bibfnamefont {G.-Q.}\ \bibnamefont {Zhang}},
		\bibinfo {author} {\bibfnamefont {D.}~\bibnamefont {Zhang}}, \bibinfo
		{author} {\bibfnamefont {T.-F.}\ \bibnamefont {Li}}, \bibinfo {author}
		{\bibfnamefont {C.-M.}\ \bibnamefont {Hu}}, \ and\ \bibinfo {author}
		{\bibfnamefont {J.~Q.}\ \bibnamefont {You}},\ }\href {\doibase
		10.1103/PhysRevLett.120.057202} {\bibfield  {journal} {\bibinfo  {journal}
			{Phys. Rev. Lett.}\ }\textbf {\bibinfo {volume} {120}},\ \bibinfo {pages}
		{057202} (\bibinfo {year} {2018})}\BibitemShut {NoStop}%
	\bibitem [{\citenamefont {Li}\ \emph {et~al.}(2018)\citenamefont {Li},
		\citenamefont {Zhu},\ and\ \citenamefont {Agarwal}}]{PhysRevLett.121.203601}%
	\BibitemOpen
	\bibfield  {author} {\bibinfo {author} {\bibfnamefont {J.}~\bibnamefont
			{Li}}, \bibinfo {author} {\bibfnamefont {S.-Y.}\ \bibnamefont {Zhu}}, \ and\
		\bibinfo {author} {\bibfnamefont {G.~S.}\ \bibnamefont {Agarwal}},\ }\href
	{\doibase 10.1103/PhysRevLett.121.203601} {\bibfield  {journal} {\bibinfo
			{journal} {Phys. Rev. Lett.}\ }\textbf {\bibinfo {volume} {121}},\ \bibinfo
		{pages} {203601} (\bibinfo {year} {2018})}\BibitemShut {NoStop}%
	\bibitem [{\citenamefont {Grigoryan}\ \emph {et~al.}(2018)\citenamefont
		{Grigoryan}, \citenamefont {Shen},\ and\ \citenamefont
		{Xia}}]{PhysRevB.98.024406}%
	\BibitemOpen
	\bibfield  {author} {\bibinfo {author} {\bibfnamefont {V.~L.}\ \bibnamefont
			{Grigoryan}}, \bibinfo {author} {\bibfnamefont {K.}~\bibnamefont {Shen}}, \
		and\ \bibinfo {author} {\bibfnamefont {K.}~\bibnamefont {Xia}},\ }\href
	{\doibase 10.1103/PhysRevB.98.024406} {\bibfield  {journal} {\bibinfo
			{journal} {Phys. Rev. B}\ }\textbf {\bibinfo {volume} {98}},\ \bibinfo
		{pages} {024406} (\bibinfo {year} {2018})}\BibitemShut {NoStop}%
	\bibitem [{\citenamefont {Harder}\ \emph {et~al.}(2018)\citenamefont {Harder},
		\citenamefont {Yang}, \citenamefont {Yao}, \citenamefont {Yu}, \citenamefont
		{Rao}, \citenamefont {Gui}, \citenamefont {Stamps},\ and\ \citenamefont
		{Hu}}]{PhysRevLett.121.137203}%
	\BibitemOpen
	\bibfield  {author} {\bibinfo {author} {\bibfnamefont {M.}~\bibnamefont
			{Harder}}, \bibinfo {author} {\bibfnamefont {Y.}~\bibnamefont {Yang}},
		\bibinfo {author} {\bibfnamefont {B.~M.}\ \bibnamefont {Yao}}, \bibinfo
		{author} {\bibfnamefont {C.~H.}\ \bibnamefont {Yu}}, \bibinfo {author}
		{\bibfnamefont {J.~W.}\ \bibnamefont {Rao}}, \bibinfo {author} {\bibfnamefont
			{Y.~S.}\ \bibnamefont {Gui}}, \bibinfo {author} {\bibfnamefont {R.~L.}\
			\bibnamefont {Stamps}}, \ and\ \bibinfo {author} {\bibfnamefont {C.-M.}\
			\bibnamefont {Hu}},\ }\href {\doibase 10.1103/PhysRevLett.121.137203}
	{\bibfield  {journal} {\bibinfo  {journal} {Phys. Rev. Lett.}\ }\textbf
		{\bibinfo {volume} {121}},\ \bibinfo {pages} {137203} (\bibinfo {year}
		{2018})}\BibitemShut {NoStop}%
	\bibitem [{\citenamefont {Wang}\ \emph {et~al.}(2019)\citenamefont {Wang},
		\citenamefont {Rao}, \citenamefont {Yang}, \citenamefont {Xu}, \citenamefont
		{Gui}, \citenamefont {Yao}, \citenamefont {You},\ and\ \citenamefont
		{Hu}}]{PhysRevLett.123.127202}%
	\BibitemOpen
	\bibfield  {author} {\bibinfo {author} {\bibfnamefont {Y.-P.}\ \bibnamefont
			{Wang}}, \bibinfo {author} {\bibfnamefont {J.~W.}\ \bibnamefont {Rao}},
		\bibinfo {author} {\bibfnamefont {Y.}~\bibnamefont {Yang}}, \bibinfo {author}
		{\bibfnamefont {P.-C.}\ \bibnamefont {Xu}}, \bibinfo {author} {\bibfnamefont
			{Y.~S.}\ \bibnamefont {Gui}}, \bibinfo {author} {\bibfnamefont {B.~M.}\
			\bibnamefont {Yao}}, \bibinfo {author} {\bibfnamefont {J.~Q.}\ \bibnamefont
			{You}}, \ and\ \bibinfo {author} {\bibfnamefont {C.-M.}\ \bibnamefont {Hu}},\
	}\href {\doibase 10.1103/PhysRevLett.123.127202} {\bibfield  {journal}
		{\bibinfo  {journal} {Phys. Rev. Lett.}\ }\textbf {\bibinfo {volume} {123}},\
		\bibinfo {pages} {127202} (\bibinfo {year} {2019})}\BibitemShut {NoStop}%
	\bibitem [{\citenamefont {Yu}\ \emph {et~al.}(2019{\natexlab{b}})\citenamefont
		{Yu}, \citenamefont {Wang}, \citenamefont {Yuan},\ and\ \citenamefont
		{Xiao}}]{PhysRevLett.123.227201}%
	\BibitemOpen
	\bibfield  {author} {\bibinfo {author} {\bibfnamefont {W.}~\bibnamefont
			{Yu}}, \bibinfo {author} {\bibfnamefont {J.}~\bibnamefont {Wang}}, \bibinfo
		{author} {\bibfnamefont {H.~Y.}\ \bibnamefont {Yuan}}, \ and\ \bibinfo
		{author} {\bibfnamefont {J.}~\bibnamefont {Xiao}},\ }\href {\doibase
		10.1103/PhysRevLett.123.227201} {\bibfield  {journal} {\bibinfo  {journal}
			{Phys. Rev. Lett.}\ }\textbf {\bibinfo {volume} {123}},\ \bibinfo {pages}
		{227201} (\bibinfo {year} {2019}{\natexlab{b}})}\BibitemShut {NoStop}%
	\bibitem [{\citenamefont {Cao}\ and\ \citenamefont
		{Yan}(2019)}]{PhysRevB.99.214415}%
	\BibitemOpen
	\bibfield  {author} {\bibinfo {author} {\bibfnamefont {Y.}~\bibnamefont
			{Cao}}\ and\ \bibinfo {author} {\bibfnamefont {P.}~\bibnamefont {Yan}},\
	}\href {\doibase 10.1103/PhysRevB.99.214415} {\bibfield  {journal} {\bibinfo
			{journal} {Phys. Rev. B}\ }\textbf {\bibinfo {volume} {99}},\ \bibinfo
		{pages} {214415} (\bibinfo {year} {2019})}\BibitemShut {NoStop}%
	\bibitem [{\citenamefont {Lachance-Quirion}\ \emph {et~al.}(2020)\citenamefont
		{Lachance-Quirion}, \citenamefont {Wolski}, \citenamefont {Tabuchi},
		\citenamefont {Kono}, \citenamefont {Usami},\ and\ \citenamefont
		{Nakamura}}]{Lachance-Quirion425}%
	\BibitemOpen
	\bibfield  {author} {\bibinfo {author} {\bibfnamefont {D.}~\bibnamefont
			{Lachance-Quirion}}, \bibinfo {author} {\bibfnamefont {S.~P.}\ \bibnamefont
			{Wolski}}, \bibinfo {author} {\bibfnamefont {Y.}~\bibnamefont {Tabuchi}},
		\bibinfo {author} {\bibfnamefont {S.}~\bibnamefont {Kono}}, \bibinfo {author}
		{\bibfnamefont {K.}~\bibnamefont {Usami}}, \ and\ \bibinfo {author}
		{\bibfnamefont {Y.}~\bibnamefont {Nakamura}},\ }\href {\doibase
		10.1126/science.aaz9236} {\bibfield  {journal} {\bibinfo  {journal}
			{Science}\ }\textbf {\bibinfo {volume} {367}},\ \bibinfo {pages} {425}
		(\bibinfo {year} {2020})}\BibitemShut {NoStop}%
	\bibitem [{\citenamefont {Zhao}\ \emph {et~al.}(2020)\citenamefont {Zhao},
		\citenamefont {Liu}, \citenamefont {Wu}, \citenamefont {Duan}, \citenamefont
		{Liu},\ and\ \citenamefont {Du}}]{PhysRevApplied.13.014053}%
	\BibitemOpen
	\bibfield  {author} {\bibinfo {author} {\bibfnamefont {J.}~\bibnamefont
			{Zhao}}, \bibinfo {author} {\bibfnamefont {Y.}~\bibnamefont {Liu}}, \bibinfo
		{author} {\bibfnamefont {L.}~\bibnamefont {Wu}}, \bibinfo {author}
		{\bibfnamefont {C.-K.}\ \bibnamefont {Duan}}, \bibinfo {author}
		{\bibfnamefont {Y.-x.}\ \bibnamefont {Liu}}, \ and\ \bibinfo {author}
		{\bibfnamefont {J.}~\bibnamefont {Du}},\ }\href {\doibase
		10.1103/PhysRevApplied.13.014053} {\bibfield  {journal} {\bibinfo  {journal}
			{Phys. Rev. Applied}\ }\textbf {\bibinfo {volume} {13}},\ \bibinfo {pages}
		{014053} (\bibinfo {year} {2020})}\BibitemShut {NoStop}%
	\bibitem [{\citenamefont {Yang}\ \emph {et~al.}(2020)\citenamefont {Yang},
		\citenamefont {Wang}, \citenamefont {Rao}, \citenamefont {Gui}, \citenamefont
		{Yao}, \citenamefont {Lu},\ and\ \citenamefont
		{Hu}}]{PhysRevLett.125.147202}%
	\BibitemOpen
	\bibfield  {author} {\bibinfo {author} {\bibfnamefont {Y.}~\bibnamefont
			{Yang}}, \bibinfo {author} {\bibfnamefont {Y.-P.}\ \bibnamefont {Wang}},
		\bibinfo {author} {\bibfnamefont {J.~W.}\ \bibnamefont {Rao}}, \bibinfo
		{author} {\bibfnamefont {Y.~S.}\ \bibnamefont {Gui}}, \bibinfo {author}
		{\bibfnamefont {B.~M.}\ \bibnamefont {Yao}}, \bibinfo {author} {\bibfnamefont
			{W.}~\bibnamefont {Lu}}, \ and\ \bibinfo {author} {\bibfnamefont {C.-M.}\
			\bibnamefont {Hu}},\ }\href {\doibase 10.1103/PhysRevLett.125.147202}
	{\bibfield  {journal} {\bibinfo  {journal} {Phys. Rev. Lett.}\ }\textbf
		{\bibinfo {volume} {125}},\ \bibinfo {pages} {147202} (\bibinfo {year}
		{2020})}\BibitemShut {NoStop}%
	\bibitem [{\citenamefont {Yu}\ \emph {et~al.}(2020{\natexlab{a}})\citenamefont
		{Yu}, \citenamefont {Shen},\ and\ \citenamefont
		{Li}}]{PhysRevLett.124.213604}%
	\BibitemOpen
	\bibfield  {author} {\bibinfo {author} {\bibfnamefont {M.}~\bibnamefont
			{Yu}}, \bibinfo {author} {\bibfnamefont {H.}~\bibnamefont {Shen}}, \ and\
		\bibinfo {author} {\bibfnamefont {J.}~\bibnamefont {Li}},\ }\href {\doibase
		10.1103/PhysRevLett.124.213604} {\bibfield  {journal} {\bibinfo  {journal}
			{Phys. Rev. Lett.}\ }\textbf {\bibinfo {volume} {124}},\ \bibinfo {pages}
		{213604} (\bibinfo {year} {2020}{\natexlab{a}})}\BibitemShut {NoStop}%
	\bibitem [{\citenamefont {Yu}\ \emph {et~al.}(2020{\natexlab{b}})\citenamefont
		{Yu}, \citenamefont {Yu},\ and\ \citenamefont {Bauer}}]{PhysRevB.102.064416}%
	\BibitemOpen
	\bibfield  {author} {\bibinfo {author} {\bibfnamefont {W.}~\bibnamefont
			{Yu}}, \bibinfo {author} {\bibfnamefont {T.}~\bibnamefont {Yu}}, \ and\
		\bibinfo {author} {\bibfnamefont {G.~E.~W.}\ \bibnamefont {Bauer}},\ }\href
	{\doibase 10.1103/PhysRevB.102.064416} {\bibfield  {journal} {\bibinfo
			{journal} {Phys. Rev. B}\ }\textbf {\bibinfo {volume} {102}},\ \bibinfo
		{pages} {064416} (\bibinfo {year} {2020}{\natexlab{b}})}\BibitemShut
	{NoStop}%
	\bibitem [{\citenamefont {Xu}\ \emph {et~al.}(2020)\citenamefont {Xu},
		\citenamefont {Zhong}, \citenamefont {Han}, \citenamefont {Jin},
		\citenamefont {Jiang},\ and\ \citenamefont {Zhang}}]{PhysRevLett.125.237201}%
	\BibitemOpen
	\bibfield  {author} {\bibinfo {author} {\bibfnamefont {J.}~\bibnamefont
			{Xu}}, \bibinfo {author} {\bibfnamefont {C.}~\bibnamefont {Zhong}}, \bibinfo
		{author} {\bibfnamefont {X.}~\bibnamefont {Han}}, \bibinfo {author}
		{\bibfnamefont {D.}~\bibnamefont {Jin}}, \bibinfo {author} {\bibfnamefont
			{L.}~\bibnamefont {Jiang}}, \ and\ \bibinfo {author} {\bibfnamefont
			{X.}~\bibnamefont {Zhang}},\ }\href {\doibase 10.1103/PhysRevLett.125.237201}
	{\bibfield  {journal} {\bibinfo  {journal} {Phys. Rev. Lett.}\ }\textbf
		{\bibinfo {volume} {125}},\ \bibinfo {pages} {237201} (\bibinfo {year}
		{2020})}\BibitemShut {NoStop}%
	\bibitem [{\citenamefont {Lachance-Quirion}\ \emph {et~al.}(2017)\citenamefont
		{Lachance-Quirion}, \citenamefont {Tabuchi}, \citenamefont {Ishino},
		\citenamefont {Noguchi}, \citenamefont {Ishikawa}, \citenamefont {Yamazaki},\
		and\ \citenamefont {Nakamura}}]{Lachance-Quirione1603150}%
	\BibitemOpen
	\bibfield  {author} {\bibinfo {author} {\bibfnamefont {D.}~\bibnamefont
			{Lachance-Quirion}}, \bibinfo {author} {\bibfnamefont {Y.}~\bibnamefont
			{Tabuchi}}, \bibinfo {author} {\bibfnamefont {S.}~\bibnamefont {Ishino}},
		\bibinfo {author} {\bibfnamefont {A.}~\bibnamefont {Noguchi}}, \bibinfo
		{author} {\bibfnamefont {T.}~\bibnamefont {Ishikawa}}, \bibinfo {author}
		{\bibfnamefont {R.}~\bibnamefont {Yamazaki}}, \ and\ \bibinfo {author}
		{\bibfnamefont {Y.}~\bibnamefont {Nakamura}},\ }\href {\doibase
		10.1126/sciadv.1603150} {\bibfield  {journal} {\bibinfo  {journal} {Science
				Advances}\ }\textbf {\bibinfo {volume} {3}} (\bibinfo {year} {2017})}\BibitemShut {NoStop}%
	\bibitem [{\citenamefont {Xiang}\ \emph {et~al.}(2013)\citenamefont {Xiang},
		\citenamefont {Ashhab}, \citenamefont {You},\ and\ \citenamefont
		{Nori}}]{RevModPhys.85.623}%
	\BibitemOpen
	\bibfield  {author} {\bibinfo {author} {\bibfnamefont {Z.-L.}\ \bibnamefont
			{Xiang}}, \bibinfo {author} {\bibfnamefont {S.}~\bibnamefont {Ashhab}},
		\bibinfo {author} {\bibfnamefont {J.~Q.}\ \bibnamefont {You}}, \ and\
		\bibinfo {author} {\bibfnamefont {F.}~\bibnamefont {Nori}},\ }\href {\doibase
		10.1103/RevModPhys.85.623} {\bibfield  {journal} {\bibinfo  {journal} {Rev.
				Mod. Phys.}\ }\textbf {\bibinfo {volume} {85}},\ \bibinfo {pages} {623}
		(\bibinfo {year} {2013})}\BibitemShut {NoStop}%
	\bibitem [{\citenamefont {Kurizki}\ \emph {et~al.}(2015)\citenamefont
		{Kurizki}, \citenamefont {Bertet}, \citenamefont {Kubo}, \citenamefont
		{M{\o}lmer}, \citenamefont {Petrosyan}, \citenamefont {Rabl},\ and\
		\citenamefont {Schmiedmayer}}]{Kurizki3866}%
	\BibitemOpen
	\bibfield  {author} {\bibinfo {author} {\bibfnamefont {G.}~\bibnamefont
			{Kurizki}}, \bibinfo {author} {\bibfnamefont {P.}~\bibnamefont {Bertet}},
		\bibinfo {author} {\bibfnamefont {Y.}~\bibnamefont {Kubo}}, \bibinfo {author}
		{\bibfnamefont {K.}~\bibnamefont {M{\o}lmer}}, \bibinfo {author}
		{\bibfnamefont {D.}~\bibnamefont {Petrosyan}}, \bibinfo {author}
		{\bibfnamefont {P.}~\bibnamefont {Rabl}}, \ and\ \bibinfo {author}
		{\bibfnamefont {J.}~\bibnamefont {Schmiedmayer}},\ }\href {\doibase
		10.1073/pnas.1419326112} {\bibfield  {journal} {\bibinfo  {journal}
			{Proceedings of the National Academy of Sciences}\ }\textbf {\bibinfo
			{volume} {112}},\ \bibinfo {pages} {3866} (\bibinfo {year}
		{2015})}\BibitemShut {NoStop}%
	\bibitem [{\citenamefont {Lachance-Quirion}\ \emph {et~al.}(2019)\citenamefont
		{Lachance-Quirion}, \citenamefont {Tabuchi}, \citenamefont {Gloppe},
		\citenamefont {Usami},\ and\ \citenamefont
		{Nakamura}}]{Lachance_Quirion_2019}%
	\BibitemOpen
	\bibfield  {author} {\bibinfo {author} {\bibfnamefont {D.}~\bibnamefont
			{Lachance-Quirion}}, \bibinfo {author} {\bibfnamefont {Y.}~\bibnamefont
			{Tabuchi}}, \bibinfo {author} {\bibfnamefont {A.}~\bibnamefont {Gloppe}},
		\bibinfo {author} {\bibfnamefont {K.}~\bibnamefont {Usami}}, \ and\ \bibinfo
		{author} {\bibfnamefont {Y.}~\bibnamefont {Nakamura}},\ }\href {\doibase
		10.7567/1882-0786/ab248d} {\bibfield  {journal} {\bibinfo  {journal} {Applied
				Physics Express}\ }\textbf {\bibinfo {volume} {12}},\ \bibinfo {pages}
		{070101} (\bibinfo {year} {2019})}\BibitemShut {NoStop}%
	\bibitem [{\citenamefont {Li}\ \emph {et~al.}(2020)\citenamefont {Li},
		\citenamefont {Zhang}, \citenamefont {Tyberkevych}, \citenamefont {Kwok},
		\citenamefont {Hoffmann},\ and\ \citenamefont
		{Novosad}}]{doi:10.1063/5.0020277}%
	\BibitemOpen
	\bibfield  {author} {\bibinfo {author} {\bibfnamefont {Y.}~\bibnamefont
			{Li}}, \bibinfo {author} {\bibfnamefont {W.}~\bibnamefont {Zhang}}, \bibinfo
		{author} {\bibfnamefont {V.}~\bibnamefont {Tyberkevych}}, \bibinfo {author}
		{\bibfnamefont {W.-K.}\ \bibnamefont {Kwok}}, \bibinfo {author}
		{\bibfnamefont {A.}~\bibnamefont {Hoffmann}}, \ and\ \bibinfo {author}
		{\bibfnamefont {V.}~\bibnamefont {Novosad}},\ }\href {\doibase
		10.1063/5.0020277} {\bibfield  {journal} {\bibinfo  {journal} {Journal of
				Applied Physics}\ }\textbf {\bibinfo {volume} {128}},\ \bibinfo {pages}
		{130902} (\bibinfo {year} {2020})}\BibitemShut {NoStop}%
	\bibitem [{\citenamefont {Wang}\ \emph {et~al.}(2016)\citenamefont {Wang},
		\citenamefont {Zhang}, \citenamefont {Zhang}, \citenamefont {Luo},
		\citenamefont {Xiong}, \citenamefont {Wang}, \citenamefont {Li},
		\citenamefont {Hu},\ and\ \citenamefont {You}}]{PhysRevB.94.224410}%
	\BibitemOpen
	\bibfield  {author} {\bibinfo {author} {\bibfnamefont {Y.-P.}\ \bibnamefont
			{Wang}}, \bibinfo {author} {\bibfnamefont {G.-Q.}\ \bibnamefont {Zhang}},
		\bibinfo {author} {\bibfnamefont {D.}~\bibnamefont {Zhang}}, \bibinfo
		{author} {\bibfnamefont {X.-Q.}\ \bibnamefont {Luo}}, \bibinfo {author}
		{\bibfnamefont {W.}~\bibnamefont {Xiong}}, \bibinfo {author} {\bibfnamefont
			{S.-P.}\ \bibnamefont {Wang}}, \bibinfo {author} {\bibfnamefont {T.-F.}\
			\bibnamefont {Li}}, \bibinfo {author} {\bibfnamefont {C.-M.}\ \bibnamefont
			{Hu}}, \ and\ \bibinfo {author} {\bibfnamefont {J.~Q.}\ \bibnamefont {You}},\
	}\href {\doibase 10.1103/PhysRevB.94.224410} {\bibfield  {journal} {\bibinfo
			{journal} {Phys. Rev. B}\ }\textbf {\bibinfo {volume} {94}},\ \bibinfo
		{pages} {224410} (\bibinfo {year} {2016})}\BibitemShut {NoStop}%
\bibitem{Gurevich}
A. G. Gurevich and G. A. Melkov, {\it Magnetization Oscillations and Waves} (CRC, Boca Raton, FL, 1996), pp. 50.
	\bibitem [{\citenamefont {Pisarchik}\ and\ \citenamefont
		{Feudel}(2014)}]{pisarchik2014control}%
	\BibitemOpen
	\bibfield  {author} {\bibinfo {author} {\bibfnamefont {A.~N.}\ \bibnamefont
			{Pisarchik}}\ and\ \bibinfo {author} {\bibfnamefont {U.}~\bibnamefont
			{Feudel}},\ }\href@noop {} {\bibfield  {journal} {\bibinfo  {journal}
			{Physics Reports}\ }\textbf {\bibinfo {volume} {540}},\ \bibinfo {pages}
		{167} (\bibinfo {year} {2014})}\BibitemShut {NoStop}%
	\bibitem [{\citenamefont {Gippius}\ \emph {et~al.}(2007)\citenamefont
		{Gippius}, \citenamefont {Shelykh}, \citenamefont {Solnyshkov}, \citenamefont
		{Gavrilov}, \citenamefont {Rubo}, \citenamefont {Kavokin}, \citenamefont
		{Tikhodeev},\ and\ \citenamefont {Malpuech}}]{PhysRevLett.98.236401}%
	\BibitemOpen
	\bibfield  {author} {\bibinfo {author} {\bibfnamefont {N.~A.}\ \bibnamefont
			{Gippius}}, \bibinfo {author} {\bibfnamefont {I.~A.}\ \bibnamefont
			{Shelykh}}, \bibinfo {author} {\bibfnamefont {D.~D.}\ \bibnamefont
			{Solnyshkov}}, \bibinfo {author} {\bibfnamefont {S.~S.}\ \bibnamefont
			{Gavrilov}}, \bibinfo {author} {\bibfnamefont {Y.~G.}\ \bibnamefont {Rubo}},
		\bibinfo {author} {\bibfnamefont {A.~V.}\ \bibnamefont {Kavokin}}, \bibinfo
		{author} {\bibfnamefont {S.~G.}\ \bibnamefont {Tikhodeev}}, \ and\ \bibinfo
		{author} {\bibfnamefont {G.}~\bibnamefont {Malpuech}},\ }\href {\doibase
		10.1103/PhysRevLett.98.236401} {\bibfield  {journal} {\bibinfo  {journal}
			{Phys. Rev. Lett.}\ }\textbf {\bibinfo {volume} {98}},\ \bibinfo {pages}
		{236401} (\bibinfo {year} {2007})}\BibitemShut {NoStop}%
	\bibitem [{\citenamefont {Para{\"i}so}\ \emph {et~al.}(2010)\citenamefont
		{Para{\"i}so}, \citenamefont {Wouters}, \citenamefont {L{\'e}ger},
		\citenamefont {Morier-Genoud},\ and\ \citenamefont
		{Deveaud-Pl{\'e}dran}}]{Paraso2010}%
	\BibitemOpen
	\bibfield  {author} {\bibinfo {author} {\bibfnamefont {T.~K.}\ \bibnamefont
			{Para{\"i}so}}, \bibinfo {author} {\bibfnamefont {M.}~\bibnamefont
			{Wouters}}, \bibinfo {author} {\bibfnamefont {Y.}~\bibnamefont {L{\'e}ger}},
		\bibinfo {author} {\bibfnamefont {F.}~\bibnamefont {Morier-Genoud}}, \ and\
		\bibinfo {author} {\bibfnamefont {B.}~\bibnamefont {Deveaud-Pl{\'e}dran}},\
	}\href {\doibase 10.1038/nmat2787} {\bibfield  {journal} {\bibinfo  {journal}
			{Nature Materials}\ }\textbf {\bibinfo {volume} {9}},\ \bibinfo {pages} {655}
		(\bibinfo {year} {2010})}\BibitemShut {NoStop}%
	\bibitem [{\citenamefont {Skardal}\ and\ \citenamefont
		{Arenas}(2019)}]{PhysRevLett.122.248301}%
	\BibitemOpen
	\bibfield  {author} {\bibinfo {author} {\bibfnamefont {P.~S.}\ \bibnamefont
			{Skardal}}\ and\ \bibinfo {author} {\bibfnamefont {A.}~\bibnamefont
			{Arenas}},\ }\href {\doibase 10.1103/PhysRevLett.122.248301} {\bibfield
		{journal} {\bibinfo  {journal} {Phys. Rev. Lett.}\ }\textbf {\bibinfo
			{volume} {122}},\ \bibinfo {pages} {248301} (\bibinfo {year}
		{2019})}\BibitemShut {NoStop}%
	\bibitem [{\citenamefont {Jung}\ \emph {et~al.}(2014)\citenamefont {Jung},
		\citenamefont {Butz}, \citenamefont {Marthaler}, \citenamefont {Fistul},
		\citenamefont {Lepp{\"a}kangas}, \citenamefont {Koshelets},\ and\
		\citenamefont {Ustinov}}]{Jung2014}%
	\BibitemOpen
	\bibfield  {author} {\bibinfo {author} {\bibfnamefont {P.}~\bibnamefont
			{Jung}}, \bibinfo {author} {\bibfnamefont {S.}~\bibnamefont {Butz}}, \bibinfo
		{author} {\bibfnamefont {M.}~\bibnamefont {Marthaler}}, \bibinfo {author}
		{\bibfnamefont {M.~V.}\ \bibnamefont {Fistul}}, \bibinfo {author}
		{\bibfnamefont {J.}~\bibnamefont {Lepp{\"a}kangas}}, \bibinfo {author}
		{\bibfnamefont {V.~P.}\ \bibnamefont {Koshelets}}, \ and\ \bibinfo {author}
		{\bibfnamefont {A.~V.}\ \bibnamefont {Ustinov}},\ }\href {\doibase
		10.1038/ncomms4730} {\bibfield  {journal} {\bibinfo  {journal} {Nature
				Communications}\ }\textbf {\bibinfo {volume} {5}},\ \bibinfo {pages} {3730}
		(\bibinfo {year} {2014})}\BibitemShut {NoStop}%
	\bibitem [{\citenamefont {Iniguez-Rabago}\ \emph {et~al.}(2019)\citenamefont
		{Iniguez-Rabago}, \citenamefont {Li},\ and\ \citenamefont
		{Overvelde}}]{Iniguez-Rabago2019}%
	\BibitemOpen
	\bibfield  {author} {\bibinfo {author} {\bibfnamefont {A.}~\bibnamefont
			{Iniguez-Rabago}}, \bibinfo {author} {\bibfnamefont {Y.}~\bibnamefont {Li}},
		\ and\ \bibinfo {author} {\bibfnamefont {J.~T.~B.}\ \bibnamefont
			{Overvelde}},\ }\href {\doibase 10.1038/s41467-019-13319-7} {\bibfield
		{journal} {\bibinfo  {journal} {Nature Communications}\ }\textbf {\bibinfo
			{volume} {10}},\ \bibinfo {pages} {5577} (\bibinfo {year}
		{2019})}\BibitemShut {NoStop}%
	\bibitem [{\citenamefont {Weng}\ \emph {et~al.}(2020)\citenamefont {Weng},
		\citenamefont {Bouchand},\ and\ \citenamefont
		{Kippenberg}}]{PhysRevX.10.021017}%
	\BibitemOpen
	\bibfield  {author} {\bibinfo {author} {\bibfnamefont {W.}~\bibnamefont
			{Weng}}, \bibinfo {author} {\bibfnamefont {R.}~\bibnamefont {Bouchand}}, \
		and\ \bibinfo {author} {\bibfnamefont {T.~J.}\ \bibnamefont {Kippenberg}},\
	}\href {\doibase 10.1103/PhysRevX.10.021017} {\bibfield  {journal} {\bibinfo
			{journal} {Phys. Rev. X}\ }\textbf {\bibinfo {volume} {10}},\ \bibinfo
		{pages} {021017} (\bibinfo {year} {2020})}\BibitemShut {NoStop}%
	\bibitem [{\citenamefont {Hellmann}\ \emph {et~al.}(2020)\citenamefont
		{Hellmann}, \citenamefont {Schultz}, \citenamefont {Jaros}, \citenamefont
		{Levchenko}, \citenamefont {Kapitaniak}, \citenamefont {Kurths},\ and\
		\citenamefont {Maistrenko}}]{Hellmann2020}%
	\BibitemOpen
	\bibfield  {author} {\bibinfo {author} {\bibfnamefont {F.}~\bibnamefont
			{Hellmann}}, \bibinfo {author} {\bibfnamefont {P.}~\bibnamefont {Schultz}},
		\bibinfo {author} {\bibfnamefont {P.}~\bibnamefont {Jaros}}, \bibinfo
		{author} {\bibfnamefont {R.}~\bibnamefont {Levchenko}}, \bibinfo {author}
		{\bibfnamefont {T.}~\bibnamefont {Kapitaniak}}, \bibinfo {author}
		{\bibfnamefont {J.}~\bibnamefont {Kurths}}, \ and\ \bibinfo {author}
		{\bibfnamefont {Y.}~\bibnamefont {Maistrenko}},\ }\href {\doibase
		10.1038/s41467-020-14417-7} {\bibfield  {journal} {\bibinfo  {journal}
			{Nature Communications}\ }\textbf {\bibinfo {volume} {11}},\ \bibinfo {pages}
		{592} (\bibinfo {year} {2020})}\BibitemShut {NoStop}%
	\bibitem [{\citenamefont {Nair}\ \emph {et~al.}(2020)\citenamefont {Nair},
		\citenamefont {Zhang}, \citenamefont {Scully},\ and\ \citenamefont
		{Agarwal}}]{PhysRevB.102.104415}%
	\BibitemOpen
	\bibfield  {author} {\bibinfo {author} {\bibfnamefont {J.~M.~P.}\
			\bibnamefont {Nair}}, \bibinfo {author} {\bibfnamefont {Z.}~\bibnamefont
			{Zhang}}, \bibinfo {author} {\bibfnamefont {M.~O.}\ \bibnamefont {Scully}}, \
		and\ \bibinfo {author} {\bibfnamefont {G.~S.}\ \bibnamefont {Agarwal}},\
	}\href {\doibase 10.1103/PhysRevB.102.104415} {\bibfield  {journal} {\bibinfo
			{journal} {Phys. Rev. B}\ }\textbf {\bibinfo {volume} {102}},\ \bibinfo
		{pages} {104415} (\bibinfo {year} {2020})}\BibitemShut {NoStop}%
	\bibitem [{\citenamefont {Bi}\ \emph {et~al.}(2021)\citenamefont {Bi},
		\citenamefont {Yan}, \citenamefont {Zhang},\ and\ \citenamefont
		{Xiao}}]{PhysRevB.103.104411}%
	\BibitemOpen
	\bibfield  {author} {\bibinfo {author} {\bibfnamefont {M.~X.}\ \bibnamefont
			{Bi}}, \bibinfo {author} {\bibfnamefont {X.~H.}\ \bibnamefont {Yan}},
		\bibinfo {author} {\bibfnamefont {Y.}~\bibnamefont {Zhang}}, \ and\ \bibinfo
		{author} {\bibfnamefont {Y.}~\bibnamefont {Xiao}},\ }\href {\doibase
		10.1103/PhysRevB.103.104411} {\bibfield  {journal} {\bibinfo  {journal}
			{Phys. Rev. B}\ }\textbf {\bibinfo {volume} {103}},\ \bibinfo {pages}
		{104411} (\bibinfo {year} {2021})}\BibitemShut {NoStop}%
	\bibitem [{\citenamefont {Ravelet}\ \emph {et~al.}(2004)\citenamefont
		{Ravelet}, \citenamefont {Mari\'e}, \citenamefont {Chiffaudel},\ and\
		\citenamefont {Daviaud}}]{PhysRevLett.93.164501}%
	\BibitemOpen
	\bibfield  {author} {\bibinfo {author} {\bibfnamefont {F.}~\bibnamefont
			{Ravelet}}, \bibinfo {author} {\bibfnamefont {L.}~\bibnamefont {Mari\'e}},
		\bibinfo {author} {\bibfnamefont {A.}~\bibnamefont {Chiffaudel}}, \ and\
		\bibinfo {author} {\bibfnamefont {F.~m.~c.}\ \bibnamefont {Daviaud}},\ }\href
	{\doibase 10.1103/PhysRevLett.93.164501} {\bibfield  {journal} {\bibinfo
			{journal} {Phys. Rev. Lett.}\ }\textbf {\bibinfo {volume} {93}},\ \bibinfo
		{pages} {164501} (\bibinfo {year} {2004})}\BibitemShut {NoStop}%
	\bibitem [{\citenamefont {Yang}\ \emph {et~al.}(2006)\citenamefont {Yang},
		\citenamefont {Ouyang}, \citenamefont {Ma}, \citenamefont {Tseng},\ and\
		\citenamefont {Chu}}]{https://doi.org/10.1002/adfm.200500429}%
	\BibitemOpen
	\bibfield  {author} {\bibinfo {author} {\bibfnamefont {Y.}~\bibnamefont
			{Yang}}, \bibinfo {author} {\bibfnamefont {J.}~\bibnamefont {Ouyang}},
		\bibinfo {author} {\bibfnamefont {L.}~\bibnamefont {Ma}}, \bibinfo {author}
		{\bibfnamefont {R.-H.}\ \bibnamefont {Tseng}}, \ and\ \bibinfo {author}
		{\bibfnamefont {C.-W.}\ \bibnamefont {Chu}},\ }\href {\doibase
		https://doi.org/10.1002/adfm.200500429} {\bibfield  {journal} {\bibinfo
			{journal} {Advanced Functional Materials}\ }\textbf {\bibinfo {volume}
			{16}},\ \bibinfo {pages} {1001} (\bibinfo {year} {2006})}\BibitemShut
	{NoStop}%
	\bibitem [{\citenamefont {Cerna}\ \emph {et~al.}(2013)\citenamefont {Cerna},
		\citenamefont {L{\'e}ger}, \citenamefont {Para{\"\i}so}, \citenamefont
		{Wouters}, \citenamefont {Morier-Genoud}, \citenamefont {Portella-Oberli},\
		and\ \citenamefont {Deveaud}}]{cerna2013ultrafast}%
	\BibitemOpen
	\bibfield  {author} {\bibinfo {author} {\bibfnamefont {R.}~\bibnamefont
			{Cerna}}, \bibinfo {author} {\bibfnamefont {Y.}~\bibnamefont {L{\'e}ger}},
		\bibinfo {author} {\bibfnamefont {T.~K.}\ \bibnamefont {Para{\"\i}so}},
		\bibinfo {author} {\bibfnamefont {M.}~\bibnamefont {Wouters}}, \bibinfo
		{author} {\bibfnamefont {F.}~\bibnamefont {Morier-Genoud}}, \bibinfo {author}
		{\bibfnamefont {M.~T.}\ \bibnamefont {Portella-Oberli}}, \ and\ \bibinfo
		{author} {\bibfnamefont {B.}~\bibnamefont {Deveaud}},\ }\href@noop {}
	{\bibfield  {journal} {\bibinfo  {journal} {Nature Communications}\ }\textbf
		{\bibinfo {volume} {4}},\ \bibinfo {pages} {2008} (\bibinfo {year}
		{2013})}\BibitemShut {NoStop}%
	\bibitem [{\citenamefont {Kubytskyi}\ \emph {et~al.}(2014)\citenamefont
		{Kubytskyi}, \citenamefont {Biehs},\ and\ \citenamefont
		{Ben-Abdallah}}]{PhysRevLett.113.074301}%
	\BibitemOpen
	\bibfield  {author} {\bibinfo {author} {\bibfnamefont {V.}~\bibnamefont
			{Kubytskyi}}, \bibinfo {author} {\bibfnamefont {S.-A.}\ \bibnamefont
			{Biehs}}, \ and\ \bibinfo {author} {\bibfnamefont {P.}~\bibnamefont
			{Ben-Abdallah}},\ }\href {\doibase 10.1103/PhysRevLett.113.074301} {\bibfield
		{journal} {\bibinfo  {journal} {Phys. Rev. Lett.}\ }\textbf {\bibinfo
			{volume} {113}},\ \bibinfo {pages} {074301} (\bibinfo {year}
		{2014})}\BibitemShut {NoStop}%
	\bibitem [{\citenamefont {Jo}\ \emph {et~al.}(2021)\citenamefont {Jo},
		\citenamefont {Kang},\ and\ \citenamefont
		{Cho}}]{https://doi.org/10.1002/advs.202004216}%
	\BibitemOpen
	\bibfield  {author} {\bibinfo {author} {\bibfnamefont {S.~B.}\ \bibnamefont
			{Jo}}, \bibinfo {author} {\bibfnamefont {J.}~\bibnamefont {Kang}}, \ and\
		\bibinfo {author} {\bibfnamefont {J.~H.}\ \bibnamefont {Cho}},\ }\href
	{\doibase https://doi.org/10.1002/advs.202004216} {\bibfield  {journal}
		{\bibinfo  {journal} {Advanced Science}\ }\textbf {\bibinfo {volume} {8}},\
		\bibinfo {pages} {2004216} (\bibinfo {year} {2021})}\BibitemShut {NoStop}%
	\bibitem [{\citenamefont {Kak}(2018)}]{kak2018ternary}%
	\BibitemOpen
	\bibfield  {author} {\bibinfo {author} {\bibfnamefont {S.}~\bibnamefont
			{Kak}},\ }\href@noop {} {\bibfield  {journal} {\bibinfo  {journal} {arXiv
				preprint arXiv:1807.06419}\ } (\bibinfo {year} {2018})}\BibitemShut {NoStop}%
	\bibitem [{\citenamefont {Luo}\ \emph {et~al.}(2020)\citenamefont {Luo},
		\citenamefont {Dong}, \citenamefont {Hu}, \citenamefont {Wang},\ and\
		\citenamefont {Duan}}]{luo2020mtl}%
	\BibitemOpen
	\bibfield  {author} {\bibinfo {author} {\bibfnamefont {L.}~\bibnamefont
			{Luo}}, \bibinfo {author} {\bibfnamefont {Z.}~\bibnamefont {Dong}}, \bibinfo
		{author} {\bibfnamefont {X.}~\bibnamefont {Hu}}, \bibinfo {author}
		{\bibfnamefont {L.}~\bibnamefont {Wang}}, \ and\ \bibinfo {author}
		{\bibfnamefont {S.}~\bibnamefont {Duan}},\ }\href@noop {} {\bibfield
		{journal} {\bibinfo  {journal} {International Journal of Bifurcation and
				Chaos}\ }\textbf {\bibinfo {volume} {30}},\ \bibinfo {pages} {2050222}
		(\bibinfo {year} {2020})}\BibitemShut {NoStop}%
\bibitem{Note1}
{In realistic problems, there are uncertainties and ambiguities, with things not simply black or white, and intermediate states can emerge. By utilizing the ternary device, one can record the intermediate state in only one physical element.}
\bibitem [{\citenamefont {Angerer}\ \emph {et~al.}(2017)\citenamefont
		{Angerer}, \citenamefont {Putz}, \citenamefont {Krimer}, \citenamefont
		{Astner}, \citenamefont {Zens}, \citenamefont {Glattauer}, \citenamefont
		{Streltsov}, \citenamefont {Munro}, \citenamefont {Nemoto}, \citenamefont
		{Rotter}, \citenamefont {Schmiedmayer},\
		and\ \citenamefont {Majer}}]{angerer2017ultralong}%
        \BibitemOpen
	\bibfield  {author} {\bibinfo {author} {\bibfnamefont {A.}~\bibnamefont
			{Angerer}}, \bibinfo {author} {\bibfnamefont {S.}~\bibnamefont {Putz}},
		\bibinfo {author} {\bibfnamefont {D.~O.}\ \bibnamefont {Krimer}}, \bibinfo
		{author} {\bibfnamefont {T.}~\bibnamefont {Astner}}, \bibinfo {author}
		{\bibfnamefont {M.}~\bibnamefont {Zens}}, \bibinfo {author} {\bibfnamefont
			{R.}~\bibnamefont {Glattauer}}, \bibinfo {author} {\bibfnamefont
			{K.}~\bibnamefont {Streltsov}}, \bibinfo {author} {\bibfnamefont {W.~J.}\
			\bibnamefont {Munro}}, \bibinfo {author} {\bibfnamefont {K.}~\bibnamefont
			{Nemoto}}, \bibinfo {author} {\bibfnamefont {S.}~\bibnamefont {Rotter}},
		\bibinfo {author} {\bibfnamefont {J.}~\bibnamefont {Schmiedmayer}}, and \bibinfo {author} {\bibfnamefont {J.}~\bibnamefont {Majer}},\ }\href@noop {} {\bibfield  {journal} {\bibinfo  {journal}
			{Science Advances}\ }\textbf {\bibinfo {volume} {3}},\ \bibinfo {pages}
		{e1701626} (\bibinfo {year} {2017})}\BibitemShut {NoStop}%
	 \bibitem{Suhl}
	H. Suhl, J. Phys. Chem. Solids {\bf1}, 209 (1957)
	\bibitem [{\citenamefont {Kittel}(1948)}]{PhysRev.73.155}%
	\BibitemOpen
	\bibfield  {author} {\bibinfo {author} {\bibfnamefont {C.}~\bibnamefont
			{Kittel}},\ }\href {\doibase 10.1103/PhysRev.73.155} {\bibfield  {journal}
		{\bibinfo  {journal} {Phys. Rev.}\ }\textbf {\bibinfo {volume} {73}},\
		\bibinfo {pages} {155} (\bibinfo {year} {1948})}\BibitemShut {NoStop}%
%\bibitem [{\citenamefont {Polkovnikov}\ \emph {et~al.}(2011)\citenamefont
%  {Polkovnikov}, \citenamefont {Sengupta}, \citenamefont {Silva},\ and\
%  \citenamefont {Vengalattore}}]{polkovnikov2011colloquium}%
%  \BibitemOpen
%  \bibfield  {author} {\bibinfo {author} {\bibfnamefont {A.}~\bibnamefont
%  {Polkovnikov}}, \bibinfo {author} {\bibfnamefont {K.}~\bibnamefont
%  {Sengupta}}, \bibinfo {author} {\bibfnamefont {A.}~\bibnamefont {Silva}}, \
%  and\ \bibinfo {author} {\bibfnamefont {M.}~\bibnamefont {Vengalattore}},\
%  }\href@noop {} {\bibfield  {journal} {\bibinfo  {journal} {Reviews of Modern
%  Physics}\ }\textbf {\bibinfo {volume} {83}},\ \bibinfo {pages} {863}
%  (\bibinfo {year} {2011})}\BibitemShut {NoStop}%
%\bibitem [{\citenamefont {Winful}\ \emph {et~al.}(1979)\citenamefont {Winful},
%  \citenamefont {Marburger},\ and\ \citenamefont {Garmire}}]{winful1979theory}%
%  \BibitemOpen
%  \bibfield  {author} {\bibinfo {author} {\bibfnamefont {H.~G.}\ \bibnamefont
%  {Winful}}, \bibinfo {author} {\bibfnamefont {J.}~\bibnamefont {Marburger}}, \
%  and\ \bibinfo {author} {\bibfnamefont {E.}~\bibnamefont {Garmire}},\
%  }\href@noop {} {\bibfield  {journal} {\bibinfo  {journal} {Applied Physics
%  Letters}\ }\textbf {\bibinfo {volume} {35}},\ \bibinfo {pages} {379}
%  (\bibinfo {year} {1979})}\BibitemShut {NoStop}%
   \bibitem [{sup()}]{supp}%
   \BibitemOpen
   \href@noop {} {}\bibinfo {howpublished} {See Supplemental Material at http:// for additional details of the theoretical derivations, measurement methods, data analysis, and the frequency-shift relaxation time, which includes Refs.~\cite{PhysRevLett.113.083603},\cite{PhysRevLett.113.156401},\cite{PhysRevB.94.224410},\cite{Gurevich},
   and \cite{Ref-add-1,Ref-add-2,Ref-add-3,Ref-add-4}}\BibitemShut {NoStop}%
   \bibitem{Ref-add-1}
   S. Blundell, {\it Magnetism in Condensed Matter} (Oxford University Press, Oxford, 2001).
   \bibitem{Ref-add-2}
   D. D. Stancil and A. Prabhakar, {\it Spin Waves: Theory and Applications} (Springer, New York, 2009).
   \bibitem{Ref-add-3}
   O. O. Soykal and M. E. Flatte, Phys. Rev. Lett. {\bf 104}, 077202 (2010).
   \bibitem{Ref-add-4}
   T. Holstein and H. Primakoff, Phys. Rev. {\bf 58}, 1098 (1940).
   \bibitem [{sup()}]{note2}%
   \BibitemOpen
   \href@noop {} {}\bibinfo {howpublished} {For a cosine signal with amplitude $A$, the time average of the signal square corresponds to the mean power of the signal: $\left[\int_{0}^{\frac{2\pi}{\omega}}(A\cos\omega t)^2\ dt\right]/(\frac{2\pi}{\omega})=\frac{A^2}{2}$, where $2\pi/\omega$ is the period of the signal. For two equal-amplitude cosine signals jointly supplied as the drive tone, i.e., $A \cos \omega t + A \cos (\omega t + \varphi)$, when $\varphi=2\pi/3$, the mean power of the synthetic signal is still equal to $\frac{A^2}{2}$: $\left\{\int_{0}^{\frac{2\pi}{\omega}}\left[A\cos\omega t+ A \cos (\omega t + \varphi)\right]^2\ dt\right\}/(\frac{2\pi}{\omega})=\frac{A^2}{2}$. In the experiment, this can be easily realized by adding a phase shifter at one of the input channels and carefully tuning the channel phase delay to make $\varphi=2\pi/3$}\BibitemShut {NoStop}%
\end{thebibliography}

\begin{thebibliography}{1}
	
	\bibitem{Blundell01}
	S. Blundell, \emph{Magnetism in Condensed Matter} (Oxford University Press, Oxford, 2001).
	
	\bibitem{Wang16}
	Y. P. Wang, G. Q. Zhang, D. Zhang, X. Q. Luo, W. Xiong, S. P. Wang, T. F. Li, C.-M. Hu and J. Q. You, Phys. Rev. B {\bf 94}, 224410 (2016).
	
	\bibitem{stancil2009spin}
	D. D. Stancil and A. Prabhakar, \emph{Spin Waves: Theory and Applications} (Springer, New York, 2009).
	
	\bibitem{Soykal10}
	O. O. Soykal and M. E. Flatte, Phys. Rev. Lett. {\bf 104}, 077202 (2010).
	
	\bibitem{Gurevich}
	A. G. Gurevich and G. A. Melkov,  \emph{Magnetization Oscillations and Waves} (CRC, Boca Raton, FL, 1996), pp. 50.
	
	\bibitem{Holstein40}
	T. Holstein and H. Primakoff, Phys. Rev. {\bf 58}, 1098 (1940).
	
	\bibitem{tabuchi2014hybridizing}
	Y. Tabuchi, S. Ishino, T. Ishikawa, R. Yamazaki, K. Usami, and Y. Nakamura, Phys. Rev. Lett. {\bf 113}, 083603 (2014).
	
	\bibitem{zhang2014strongly}
	X. Zhang, C.-L. Zou, L. Jiang, and H. X. Tang, Phys. Rev. Lett. {\bf 113}, 156401 (2014).
	
\end{thebibliography}
\end{document}